\documentclass[aps, prx, twocolumn, groupedaddress, notitlepage]{revtex4-1}
\usepackage{float}
\usepackage{bbm}
\usepackage{epsfig}
\usepackage{epstopdf}
\usepackage{graphicx}
\usepackage{amsmath,amssymb}
\usepackage{physics}
\usepackage{color}
\usepackage{hyperref}
\usepackage{lineno,blindtext}
\usepackage[caption=false]{subfig}
\usepackage{siunitx, booktabs}
\usepackage{diagbox, eqparbox, hhline}
\usepackage{soul}
\usepackage{xcolor}
\usepackage[normalem]{ulem}
\newcolumntype{P}[1]{>{\centering\arraybackslash}p{#1}}

\begin{document}

\title{Exploring self-consistency of the equations of axion electrodynamics in Weyl semimetals}

\author{Kuangyin Deng}
\author{John S. Van Dyke}
\author{Djordje Minic}
\author{J. J. Heremans}
\author{Edwin Barnes}
\email{efbarnes@vt.edu}
\affiliation{Department of Physics, Virginia Tech, Blacksburg, VA 24061, USA}

\begin{abstract}
Recent works have provided evidence that an axial anomaly can arise in Weyl semimetals. If this is the case, then the electromagnetic response of Weyl semimetals should be governed by the equations of axion electrodynamics. These equations capture both the chiral magnetic and anomalous Hall effects in the limit of linear response, while at higher orders their solutions can provide detectable electromagnetic signatures of the anomaly. In this work, we consider three versions of axion electrodynamics that have been proposed in the Weyl semimetal literature. These versions differ in the form of the chiral magnetic term and in whether or not the axion is treated as a dynamical field. In each case, we look for solutions to these equations for simple sample geometries subject to applied external fields. We find that in the case of a linear chiral magnetic term generated by a non-dynamical axion, self-consistent solutions can generally be obtained. In this case, the magnetic field inside of the Weyl semimetal can be magnified significantly, providing a testable signature for experiments. Self-consistent solutions can also be obtained for dynamical axions, but only in cases where the chiral magnetic term vanishes identically. Finally, for a nonlinear form of the chiral magnetic term frequently considered in the literature, we find that there are no self-consistent solutions aside from a few special cases.   
\end{abstract}

\maketitle

\section{Introduction}\label{sec:Intro}
Weyl semimetals (WSMs) have garnered substantial interest in recent years due to their topological properties and unusual transport phenomena~\cite{bera2016dirty,roy2016quantum,ramakrishnan2015transport}. While they were first theorized long ago~\cite{HerringPR1937,abrikosov1996possible}, only in recent years have explicit candidate materials been put forward and confirmed~\cite{armitage2018weyl,jia2016weyl,hasan2017discovery,ganeshan2015constructing}. They were first predicted theoretically to arise in pyrochlore iridates~\cite{wan2011topological,hosur2012charge}, and their existence was later confirmed experimentally in compounds such as TaAs and NbAs~\cite{huang2015weyl,weng2015weyl,lv2015experimental,lv2015observation,xu2015discoveryA,xu2015discoveryB}. More recently, additional WSMs have been discovered in ferromagnetic materials~\cite{LiuScience2019,MoraliScience2019,BelopolskiScience2019}. The low-energy quasiparticle excitations in WSMs are Weyl fermions, which leads to the possibility of observing interesting phenomena such as the chiral magnetic effect~\cite{FukushimaPRD2008}. These Weyl fermion quasiparticles exist near band touching points (Weyl nodes), which carry chiral topological charges. The linearly dispersing bands in the vicinity of Weyl nodes, as well as the Fermi arc states connecting node projections on the WSM surface~\cite{wan2011topological,burkov2011weyl,wang2013three}, have been observed experimentally through angle-resolved photoemission spectroscopy (ARPES)~\cite{huang2015weyl,weng2015weyl,xu2015discoveryA,xu2015discoveryB,hosur2012charge,lv2015experimental,lv2015observation}. When the Weyl nodes are close to the Fermi energy, it has been reported that electrons can achieve ultrahigh mobility~\cite{jiang2016chiral}. Other effects such as the Goos-Hänchen (GH) and Imbert-Fedorov (IF) shifts can also be produced in WSMs~\cite{jiang2015topological}. While the GH shift is valley-independent, the IF shift is valley-dependent in WSMs due to the opposite chiral charge of the Weyl nodes in momentum space. This provides an alternative way to detect Weyl node properties. Another prediction of WSMs that has drawn much attention is the axial anomaly, which can be understood to arise from the pairing of opposite chiral charges~\cite{nielsen1983adler,chen2013axion,zyuzin2012topological,vazifeh2013electromagnetic,khaidukov2018magnetostatics,goswami2013axionic,son2013chiral}. Effects of axial anomalies have previously been seen in high energy physics~\cite{adler1969axial,bell1969pcac} and in superfluids~\cite{bevan1997momentum}.

In WSMs, the axial anomaly produces two topological effects related to the Berry curvature of the Weyl nodes: the chiral magnetic effect (CME) and the anomalous Hall effect (AHE)~\cite{FukushimaPRD2008,chen2013axion,zyuzin2012topological,son2013chiral,burkov2014anomalous,goswami2013axionic,vazifeh2013electromagnetic,khaidukov2018magnetostatics}. In the CME, an external magnetic field produces a current in the same direction as the field. This effect is expected to occur in WSMs because the left and right chiral Weyl fermions become separated in energy in the presence of the external field, inducing a current referred to as the chiral magnetic current. To observe the CME experimentally, transport signatures such as a negative longitudinal magnetoresistance have been proposed and measured~\cite{huang2015observation,zhang2016signatures,li2016chiral}. However other effects, including giant magnetoresistance and large-angle scattering, can also lead to negative longitudinal magnetoresistance~\cite{he2014quantum,liang2015ultrahigh,goswami2015axial,li2016chiral,wu2016evidence,zhang2017electronic,mirlin2001quasiclassical}, making it difficult to confirm the CME in such experiments. In the AHE, an antisymmetric off-diagonal resistivity is produced from a magnetization in the sample rather than an external magnetic field \cite{nagaosa2010anomalous,ChangScience2013}. An applied electric field then generates current in a transverse direction. In general, the AHE can be rooted in the material itself (intrinsic) or arise from impurity scattering (extrinsic). In WSMs, the separation of Weyl node pairs in momentum space, combined with an axial anomaly, would cause a purely intrinsic AHE~\cite{burkov2014anomalous}. Like with negative longitudinal magnetoresistance, transport measurements showing an AHE also do not provide a unique indicator of the axial anomaly, as this effect can occur in any material that has a nonzero integral of Berry curvature~\cite{nagaosa2010anomalous}. Thus, other experimental signatures beyond transport measurements would be helpful in confirming the CME and the axial anomaly in WSMs.

Axion electrodynamics provides an alternative route for verifying the existence of the axial anomaly. If one integrates out the low-energy Weyl fermions and is left with only the electric and magnetic fields, one arrives at an effective description known as axion electrodynamics. In the literature, several approaches have been taken to derive the equations of axion electrodynamics for WSMs. In the first, one starts with a microscopic model of a WSM \cite{burkov2011weyl} and integrates out the electrons. This approach yields a non-dynamical axion field and produces a linear chiral magnetic term in Amp\`ere's law ~\cite{zyuzin2012topological,chen2013axion,goswami2013axionic}. Here, we use the term ``non-dynamical" to refer to fields that have a fixed form, while we use ``dynamical" to refer to fields whose form is determined by solving the equations of axion electrodynamics. It was subsequently found that the CME can occur in this case if time-dependent fields are applied to the WSM~\cite{vazifeh2013electromagnetic,chen2013axion}. In a second approach, one incorporates axial anomaly effects in a semiclassical Boltzmann equation~\cite{son2013chiral}. This leads to CME and AHE currents that can then be included in Maxwell's equations to produce a different form of axion electrodynamics. Here, the CME term is nonlinear in the fields and proportional to the inner product of the electric and magnetic fields ($\vec{E}\cdot\vec{B}$)~\cite{FukushimaPRD2008,li2016chiral}. In this case, a chiral current can be generated by applying time-independent external fields. Experimental observations of negative longitudinal magnetoresistance have been explained using this version of the CME term~\cite{huang2015observation,zhang2016signatures}. In the case of parallel electric and magnetic fields, similar behavior can also arise from a one-dimensional axial anomaly that generically emerges in three-dimensional metals (not necessarily WSMs) if the magnetic field is sufficiently strong~\cite{goswami2015axial}. Finally, a third approach considers chiral symmetry breaking via the formation of charge density waves in WSMs. The resulting axion insulator phase is characterized by an order parameter whose phase is a dynamical axion field~\cite{wang2013chiral}. This axion couples to the electric and magnetic fields through a topological $\theta$ term in the Maxwell action. This action yields axion electrodynamics equations that are similar to those of the first approach described above, except that now the axion is an independent dynamical field with its own equation of motion. In both the first and third approaches, the new term in the Maxwell action can also be obtained by performing a chiral transformation on the path integral measure, following the standard anomaly derivation first introduced by Fujikawa~\cite{fujikawa2004path,zyuzin2012topological,goswami2013axionic}. 

Regardless of which approach one takes to derive axion electrodynamics, one has a modified form of Maxwell's equations that govern the behavior of electric and magnetic fields in the presence of an axial anomaly. Their self-consistent solutions in the presence of applied external fields can be used to guide experiments that look for signatures of the axial anomaly. This constitutes an alternative strategy that is complementary to transport-based experiments. A first pass at this approach was taken by a subset of the authors in Ref.~\cite{barnes2016electromagnetic}. However, this earlier work neglected the AHE term altogether and did not consider dynamical axions. A full analysis of the self-consistency of the different versions of axion electrodynamics that have been put forward in the context of WSMs has yet to be carried out. It is not yet clear how the different versions relate to one another or which provides the most accurate description of a given experimental setup. These questions could also be addressed through experimental observation, provided the solutions to these equations are well understood.

In this work, we address these open questions by attempting to solve all three versions of axion electrodynamics self-consistently for simple sample geometries and various external field configurations. In the case of version 1 (non-dynamical axion, linear CME term), we solve the equations for a semi-infinite WSM slab in the presence of time-dependent, external electric and magnetic fields. We find that self-consistent solutions can generally be obtained, and that the magnetic field inside the slab can be substantially enhanced depending on the Weyl node separation and on the frequency of the applied fields. This provides a potential experimental diagnostic of the axial anomaly. For version 2 (non-dynamical axion, nonlinear CME term), we find that for a semi-infinite slab immersed in time-independent fields, self-consistent solutions generically do not exist, aside from a few special cases. We also find that while self-consistent solutions can be obtained in the case of an infinite WSM wire, the solutions always exhibit unphysical divergences along the axis of the wire. Finally, in the case of version 3 (dynamical axions, linear CME term), we show that self-consistent solutions can be obtained, but only when the CME term vanishes identically. Otherwise, the solutions violate energy conservation.

The paper is organized as follows. In Sec.~\ref{sec:ND axion: TD}, we solve the axion electrodynamics equations for non-dynamical axions in a semi-infinite slab subject to time-dependent fields. In Sec.~\ref{sec:ND axion: TID}, we consider non-dynamical axions in a semi-infinite slab, an infinite slab, and an infinite cylinder, all subject to time-independent fields. In Sec.~\ref{sec:D axion}, we generalize to the case of dynamical axions in a semi-infinite slab. We conclude in Sec.~\ref{sec:Conclu}. Several appendices contain details of the calculations summarized in Secs.~\ref{sec:ND axion: TD}-\ref{sec:D axion}.

Before moving on to our explicit solutions, we first note that throughout this work, we neglect the role of Fermi arc surface states in our analysis. One reason for this is because most of the sample geometries we focus on, namely semi-infinite slabs with the inter-Weyl node axis oriented perpendicular to the surface and cylindrically symmetric infinite wires, do not exhibit Fermi arcs. However, even in cases where Fermi arcs could arise, such as in the case of semi-infinite slabs with non-orthogonal inter-Weyl node axes, we do not expect them to significantly impact our results because their effect should be restricted to a small region close to the surface. We also note that, to our knowledge, axion electrodynamics equations that incorporate Fermi arc effects have not yet been derived.

\section{Non-dynamical axions and linear chiral magnetic term}\label{sec:ND axion: TD}

\begin{figure*}
\centering
{{\includegraphics[trim=0cm 0cm 0cm 0cm, clip=true,width=17cm, angle=0]{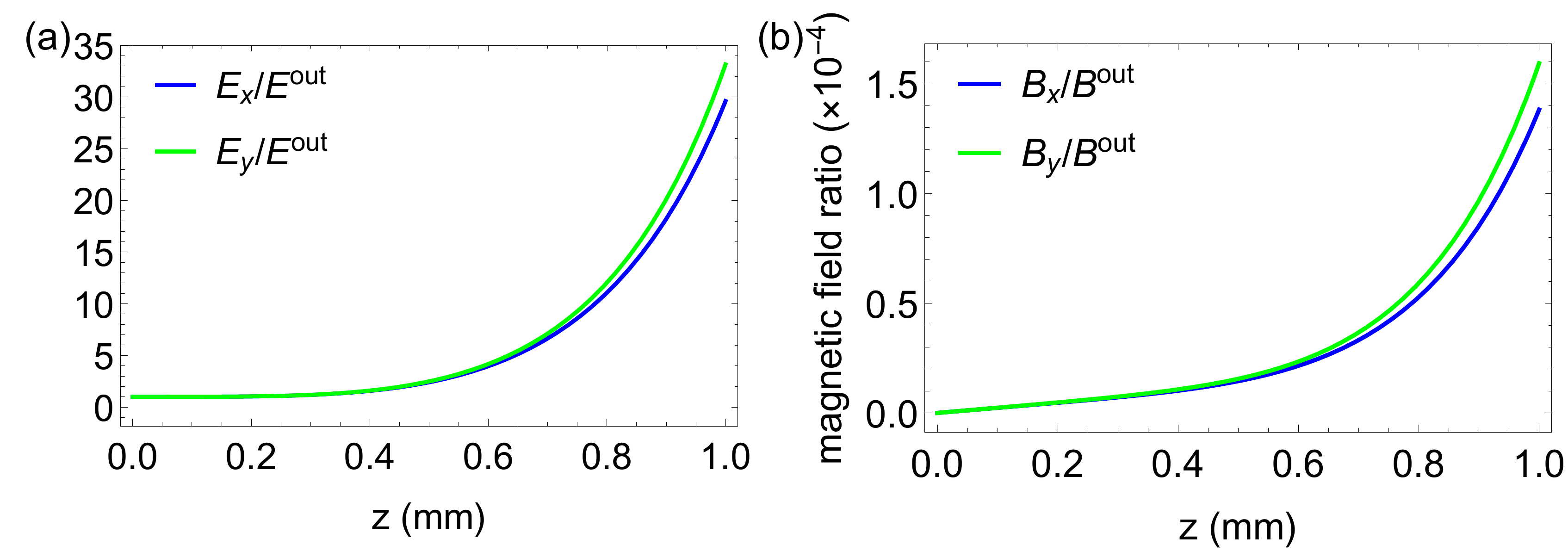}}}
\caption{(a) Electric field components and (b) magnetic field components as a function of distance $z$ inside a semi-infinite slab for $t=0$. Here we set $\Delta\varepsilon=6$ meV, $\Delta p_z c=9.873\cross 10^4$ meV, $\omega=3.0$ GHz. The boundary conditions are $E_x(0)=E_y(0)=E^{out}$, $\partial_zE_x(0)=\partial_zE_y(0)=0$ and $B^{out}=\frac{E^{out}}{c}$. The parameters in Eq.~\eqref{eq:main_generalsolution} are $d_1=-4831.62-70.95i$, $d_2=-4761.73i$, $d_3=4903.63i$ and $d_4=4831.62-70.95i$ in units of m$^{-1}$. Here $d_4$ is the only root with a positive real value, and this gives rise to the exponential growth of the fields with $z$ in this example.}\label{Fig:1}
\end{figure*}
The axial anomaly was first proposed theoretically in the context of high-energy physics~\cite{adler1969axial,bell1969pcac}. Its presence leads to an additional term $\mathcal{L}_{\theta_1}$ in the Lagrangian density:
\begin{align}
\mathcal{L}_0&=-\frac{1}{4\mu_0}F^{\alpha\beta}F_{\alpha\beta}-A_\alpha J^\alpha,\label{eq:L0}\\
\mathcal{L}_{\theta_1}&=-\frac{\kappa}{4}\theta F_{\alpha\beta}\frac{1}{2}\epsilon^{\alpha\beta\gamma\lambda}F_{\gamma\lambda}=\frac{\kappa}{c}\theta\vec{E}\cdot\vec{B},\label{eq:L1}
\end{align}
where $\mathcal{L}_0$ is the Lagrangian density for the original electromagnetic fields, and $\mathcal{L}_{\theta_1}$ is the term that describes the axion-electromagnetic interaction. $A_\alpha$ is the vector potential, while $J^\alpha$ is the source current. The signature of the metric $\eta^{\mu\nu}$ is $(1,-1,-1,-1)$,
the field strength is $F_{\alpha\beta}=\partial_\alpha A_\beta-\partial_\beta A_\alpha$, $\mu_0$ is the vacuum permeability, $c$ is the speed of light, and $\kappa$ is the coupling constant between the axion field $\theta$ and the electromagnetic field. We set $\kappa=\frac{e^2 c}{2\pi^2\hbar^2}$ following Ref.~\cite{zyuzin2012topological}. The corresponding Euler-Lagrange equations yield the first version of axion electrodynamics we consider in this work~\cite{wilczek1987two,sikivie1983experimental}:
\begin{align}
\vec{\nabla}\cdot\vec{E}&=\frac{\rho}{\varepsilon_0}-\mu_0c\kappa\vec{\nabla}\theta\cdot\vec{B},\label{eq: derivation E1}\\
\vec{\nabla}\cross\vec{E}&=-\frac{\partial\vec{B}}{\partial t},\label{eq: derivation E2}\\
\vec{\nabla}\cdot\vec{B}&=0,\label{eq: derivation B1}\\
\vec{\nabla}\cross\vec{B}&=\frac{1}{c^2}\frac{\partial\vec{E}}{\partial t}+\mu_0\vec{j}+\frac{\mu_0\kappa}{c}(\partial_t\theta\vec{B}+\vec{\nabla}\theta\cross\vec{E})\label{eq: derivation B2},
\end{align}
where $c^2=\frac{1}{\mu_0\varepsilon_0}$. In WSMs, effective axions form due to linear band crossings, creating Weyl fermions with definite chiralities. Ref.~\cite{zyuzin2012topological} obtained the following expression for the axion field $\theta$ for WSMs using Fujikawa's method~\cite{fujikawa2004path}:
\begin{align}
\theta(\vec{r},t)=\Delta\vec{p}\cdot\vec{r}-\Delta\varepsilon t,\label{eq: theta form}
\end{align}
where $\Delta\vec{p}$ and $\Delta\varepsilon$ are the momentum and energy separation of a pair of Weyl nodes, respectively. Here, we have defined the coordinates $(\vec{r},t)=(x,y,z,t)$. Although the axion field $\theta(\vec{r},t)$ itself depends on the choice of coordinate origin, this choice does not affect the solutions of the axion electrodynamics equations since only derivatives of $\theta(\vec{r},t)$ enter into these equations. As shown in Ref.~\cite{kargarian2015theory}, the AHE term $\Delta\vec{p}\cdot\vec{r}$ can lead to interesting electromagnetic responses such as Kerr and Faraday rotations. For simplicity, here we focus on materials with a single pair of Weyl nodes separated in both momentum and energy, as can occur in WSMs with broken time-reversal symmetry~\cite{LiuScience2019,MoraliScience2019,BelopolskiScience2019}. Multiple Weyl node pairs near the Fermi surface would lead to a linear superposition of $\theta$-dependent terms (one term for each node pair) in Eqs.~\eqref{eq: derivation E1} and \eqref{eq: derivation B2}, which would effectively modify the coefficients multiplying the electromagnetic fields in these terms but otherwise leave the axion equations intact. In order for the CME term---the term proportional to $\partial_t\theta$ in Eq.~\eqref{eq: derivation B2}---to be present in these equations, the electric and magnetic fields have to be time-dependent~\cite{vazifeh2013electromagnetic,chen2013axion}. Using the same coordinates defined above, we set $\vec{E}(\vec{r},t)=e^{i\omega t}\vec{E}(\vec{r})$, $\vec{B}(\vec{r},t)=e^{i\omega t}\vec{B}(\vec{r})$. Hence the system is driven by a single frequency, and the spatial part can be separated from the time-dependent part for the electromagnetic fields. For the current $\vec{j}$, we implement Ohm's law,
\begin{align}
\vec{j}=\sigma_0\vec{E}.
\end{align}
In principle, the conductivity $\sigma$ is frequency and temperature dependent (calculated by Ref.~\cite{throckmorton2015many}):
\begin{align}
\sigma(\omega)=\frac{1}{i\omega+\frac{1}{\tau}}\cross\frac{v_F^2e^2g}{3\pi^2(\hbar v_F)^3}\int_0^\infty {d\varepsilon \varepsilon^2(-\frac{\partial f^0(\varepsilon,T)}{\partial\varepsilon})}.
\end{align}
Here $e$, $v_F$, $g$ and $\tau$ are the electron charge, Fermi velocity, light-matter coupling and scattering time, respectively. $f^0(\varepsilon,T)$ is the Fermi-Dirac distribution. The integral above leads to a constant decided by the temperature. Considering the limit $\omega\rightarrow 0$ and $T\rightarrow 0$, denoting $\sigma_0=\sigma(0)$, we have (see App.~\ref{app:cond})
\begin{align}
\sigma_0=\frac{e^2g\tau k_F^2v_F}{3\pi^2\hbar^3},\label{eq:cond}
\end{align}
whereas the carrier density is $n=gk_F^3/6\pi$. Thus we have the relation $\sigma_0\propto n^{\frac{2}{3}}$. In WSMs, $n$ is typically very low since $k_F$ is small around Weyl nodes. When this happens, the Ohmic conductivity can be ignored, and we can set $\vec{j}=0$ in the axion equations. If $n$ is increased sufficiently (e.g., through doping), at some point the conductivity can no longer be ignored, and the current cannot be set to zero. Below, we consider each of these two cases separately. In both cases, we consider a semi-infinite slab where the WSM fills the half-space $z\geq0$. By symmetry, the fields can only depend on the $z$ coordinate. Furthermore, the relaxation time $\tau$ in Eq.~\eqref{eq:cond} is related to the Weyl separations~\cite{li2016chiral} since the scattering happens between the Weyl nodes. However, this does not affect the fact that $\sigma_0$ is a constant spatially.

\subsection{Zero current case: $\vec{j}=0$}
We first consider the case where the electron density is very low, so that $\sigma_0$ is small, and we can set $\vec{j}=0$. We show in Appendix~\ref{app:NDTD1} that Eqs.~\eqref{eq: derivation E1}-\eqref{eq: derivation B2} reduce to the following set of equations governing the fields inside the WSM:
\begin{align}
&\partial_z^2E_y+\frac{\omega^2}{c^2}E_y-\frac{\mu_0\kappa}{c}\Delta\varepsilon\partial_zE_x-i\omega\frac{\mu_0\kappa}{c}\Delta p_zE_x\nonumber\\
&-\mu_0^2\kappa^2\Delta p_x(\Delta p_xE_y-\Delta p_yE_x)=0, \label{eq:main_reduced1}\\
&\partial_z^2E_x+\frac{\omega^2}{c^2}E_x+\frac{\mu_0\kappa}{c}\Delta\varepsilon\partial_zE_y+i\omega\frac{\mu_0\kappa}{c}\Delta p_zE_y\nonumber\\
&+\mu_0^2\kappa^2\Delta p_y(\Delta p_xE_y-\Delta p_yE_x)=0. \label{eq:main_reduced2}
\end{align}
\begin{align}
E_z=\mu_0\kappa c\frac{i}{\omega}(\Delta p_xE_y-\Delta p_yE_x).\label{eq:TDS_B4_3}
\end{align}
\begin{equation}
B_z=0,\quad
B_y=\frac{i}{\omega}\partial_z E_x,\quad
B_x=-\frac{i}{\omega}\partial_z E_y.
\end{equation}
The general solutions to Eqs.~\eqref{eq:main_reduced1} and \eqref{eq:main_reduced2} have the form
\begin{equation}
     E_y=\sum_{i=1}^4a_ie^{d_iz},\qquad
     E_x=\sum_{i=1}^4b_ie^{d_iz},
\label{eq:main_generalsolution}
\end{equation}
where the parameters $d_i$ depend on the frequency $\omega$ of the applied fields and on the energy and momentum separations of the Weyl nodes, $\Delta\varepsilon$ and $\Delta\vec p$. The $d_i$ are the roots of a characteristic equation
whose explicit form is given in Appendix~\ref{app:NDTD1}. The remaining 8 coefficients, $a_i$ and $b_i$, are determined by Eqs.~\eqref{eq:main_reduced1} and \eqref{eq:main_reduced2} and by the boundary conditions. We show in the Appendix that the fields are always continuous at the surface of the WSM. In general, we find self-consistent solutions for any choice of the applied external fields.

\begin{figure*}[!tbp]
\centering
{{\includegraphics[trim=0cm 0cm 0cm 0cm, clip=true,width=18cm, angle=0]{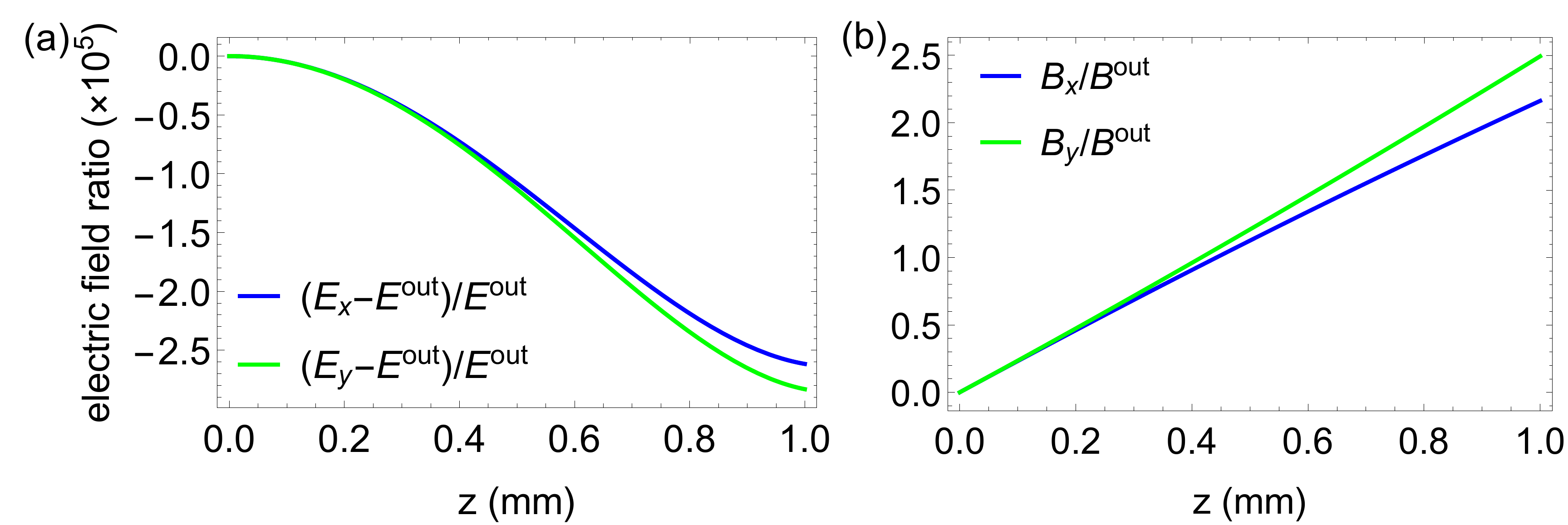}}}
\caption{(a) Electric field components and (b) magnetic field components as a function of distance $z$ inside a semi-infinite slab for $t=0$. Here we set $\Delta\varepsilon=6$ meV, $\Delta p_z c=98.73$ meV, $\omega=3.0$ GHz. The boundary conditions are $E_x(0)=E_y(0)=E^{out}$, $\partial_zE_x(0)=\partial_zE_y(0)=0$ and $B^{out}=\frac{E^{out}}{c}$. The parameters in Eq.~\eqref{eq:main_generalsolution} are $d_1=-134.966 - 70.95i$, $d_2=-97.8207i$, $d_3=239.721i$ and $d_4=134.966 - 70.95i$ in units of m$^{-1}$. Here, $d_4$ is the only root with a positive real value, and this causes the slow growth of the fields with increasing $z$.}\label{Fig:2}
\end{figure*}

As an explicit example, consider the case where the Weyl node momentum separation is in the $z$ direction, i.e., $\Delta p_x=\Delta p_y=0$ and $\Delta p_z \neq 0$. In this case, the characteristic equation is (see Appendix~\ref{app:NDTD1}):
\begin{align}
&d^4+(2\frac{\omega^2}{c^2}+\Delta\varepsilon^2\frac{\mu_0^2\kappa^2}{c^2})d^2+2i\omega\Delta\varepsilon
\frac{\mu_0^2\kappa^2}{c^2}\Delta p_zd\nonumber\\ &+\frac{\omega^4}{c^4}-\frac{\mu_0^2\kappa^2\omega^2}{c^2}\Delta p_z^2=0.
\end{align}
The energy and momentum separations of Weyl nodes are typically on the order of $\Delta\varepsilon\sim 1$ meV to $20$ meV~\cite{xu2015discoveryA} and $\Delta k\sim 0.05$~{\AA}$^{-1}$~\cite{hasan2017discovery,belopolski2016discovery,xu2015discoveryA}, respectively. Note that the energy separation can arise as a consequence of breaking both inversion and time-reversal symmetry, as discussed theoretically in Ref.~\cite{zyuzin2012weyl,burkov2011weyl,murakami2007phase}. This can occur for example in noncentrosymmetric and ferromagnetic WSMs, as predicted by first-principles studies~\cite{chang2018magnetic}. Alternatively, one can start with a noncentrosymmetric compound and apply a static magnetic field to break time-reversal symmetry~\cite{cano2017chiral}. Based on these possibilities, we make the following parameter choices: $\Delta\varepsilon=6$ meV, $\Delta p_z c=\hbar \Delta k_z c=9.873\cross 10^4$ meV, 
and we set the frequency to $\omega=3.0$ GHz. We take the fields outside the WSM ($z<0$) to be
\begin{align}
E_x=E_y&=E^{out}\cos{\frac{\omega}{c}z},\label{eq:main_TDS_vac_E1}\\
B_x=-B_y&=\frac{i}{c}E^{out}\sin{\frac{\omega}{c}z}\label{eq:main_TDS_vac_B2}.
\end{align} 
This choice then implies the following boundary conditions for the fields inside the slab: $E_x(0)=E_y(0)=E^{out}$ and $\partial_zE_x(0)=\partial_zE_y(0)=0$. The resulting electric and magnetic fields inside the WSM for these parameters at $t=0$ are shown in Fig.~\hyperref[Fig:1]{\ref*{Fig:1}}. We see that both fields increase quickly with depth $z$ into the slab, providing a detectable signature of the anomaly. The fields also oscillate, but the oscillation period is very long, approximately 89~mm for the parameters chosen in this example. This value is determined by the $d_i$, the precise values of which are quoted in the figure caption. It is also evident in Fig.~\hyperref[Fig:1]{\ref*{Fig:1}(b)} that the magnetic field grows particularly fast with increasing $z$, reaching an amplitude that is approximately $1.5\times 10^4$ larger than the magnetic field outside the WSM at a depth of $z=1$ mm. This rapid growth must ultimately saturate at a maximal value in a real sample, perhaps due to impurity scattering or other effects not accounted for here.

The magnification of the magnetic field inside the slab is due to the fact that the momentum separation between the Weyl nodes is much larger than their energy separation. If we reduce the momentum separation by a factor of $10^3$ ($\Delta p_z c=98.73$ meV), which is still significantly larger than the energy separation (keeping other parameters fixed), we obtain the results in Fig.~\hyperref[Fig:2]{\ref*{Fig:2}} at $t=0$. Here we see that the amplitude of the magnetic field still increases with $z$, but now only reaches about $2.5$ times the applied field at $z=1$~mm. Note that the momentum separation between Weyl nodes is in principle adjustable using an applied magnetic field~\cite{gorbarPRB2013,cano2017chiral}, making it possible to probe this transition in behavior. We generally find exponentially growing solutions like those shown in Figs.~\hyperref[Fig:1]{\ref*{Fig:1}} and \hyperref[Fig:2]{\ref*{Fig:2}} when the frequency $\omega$ is higher than $10^5$ Hz. However, when $\Delta p_zc$ is on the order of the energy separation, for example $\Delta p_zc=9.873$ meV, the field amplitudes inside can be 3 or more orders of magnitude smaller than those of the applied fields ($E\sim1.5E^{out}$ and $B\sim 0.0004B^{out}$) and the solutions become purely oscillatory with strictly imaginary $d_i$ rather than exponentially growing. In this regime, the oscillation period is in the range 10 - 200 mm. In addition to decreasing the Weyl node momentum separation, one can also lower the frequency of the applied fields to get oscillatory solutions. When $\omega\lesssim10^5$ Hz, all the $d_i$ become purely imaginary even if $\Delta p_z$ remains large (e.g, $\Delta p_zc=9.873\cross 10^4$ meV), in which case the fields inside are purely oscillatory. In this case, the maximal amplitudes of the fields inside are comparable to those outside the WSM ($E\sim1.5E^{out}$ and $B\sim 4B^{out}$). The oscillation period remains in the range of 10 - 200~mm in this case.

One might worry about whether energy is conserved in our solutions in light of the substantial magnification of the magnetic field inside the slab that occurs for $\omega\gtrsim10^5$~Hz. On each side of the boundary, the energy and momentum are conserved if the energy-momentum tensor obeys the equations
$\partial_\mu T^{\mu\nu}=0$.
This is automatically satisfied if $T^{\mu\nu}$ is derived from the Lagrangian and if we assume the energy density is continuous across the boundary. We show the explicit form of the energy-momentum tensor in Sec.~\ref{sec:D axion}, where we find that the energy density is continuous across the boundary provided we choose the right boundary conditions for the axion field. With this consideration in mind, we conclude that this version of axion electrodynamics (with a fixed background axion field and a linear chiral magnetic term) generally has self-consistent solutions.

\subsection{Non-zero current case: $\vec{j}\ne0$}
Next, we consider the case where the electron density is sufficiently large that the Ohmic current cannot be neglected. Adapting the same form for the electromagnetic fields as before, $\vec{E}(\vec{r},t)=e^{i\omega t}\vec{E}(\vec{r})$ and $\vec{B}(\vec{r},t)=e^{i\omega t}\vec{B}(\vec{r})$, and using Ohm's law $\vec{j}=\sigma_0\vec{E}$, we can write the current in a separated form as well: $\vec{j}(\vec{r},t)=e^{i\omega t}\vec{j}(\vec{r})$. When the conductivity is nonzero, the charge density $\rho$ must also be nonzero unless $\Delta p_x=\Delta p_x=0$, as we show in Appendix~\ref{app:NDTD2}.

For simplicity, we consider the solutions under the assumption $\Delta p_x=\Delta p_y=0$, as in the previous subsection. In this case we have $E_z=0$ (see Appendix~\ref{app:NDTD2}). We also show in Appendix~\ref{app:NDTD2} that Eqs.~\eqref{eq: derivation E1}-\eqref{eq: derivation B2} reduce to the following set of equations governing the fields inside the WSM:
\begin{align}
\partial_z^2E_y+\frac{\omega^2}{c^2}E_y-i\mu_0\sigma_0\omega E_y\nonumber\\
-\frac{\mu_0\kappa}{c}\Delta\varepsilon\partial_zE_x-i\omega\frac{\mu_0\kappa}{c}\Delta p_zE_x&=0,\\
\partial_z^2E_x+\frac{\omega^2}{c^2}E_x-i\mu_0\sigma_0\omega E_x\nonumber\\
+\frac{\mu_0\kappa}{c}\Delta\varepsilon\partial_zE_y+i\omega\frac{\mu_0\kappa}{c}\Delta p_zE_y&=0.
\end{align}
Similarly to before, the operator equation becomes
\begin{align}\label{eq_eqn_for_d}
&d^4+[2(\frac{\omega^2}{c^2}-i\mu_0\sigma_0\omega)+\Delta\varepsilon^2\frac{\mu_0^2\kappa^2}{c^2}]d^2+2i\omega\Delta\varepsilon
\frac{\mu_0^2\kappa^2}{c^2}\Delta p_zd\nonumber\\ &+(\frac{\omega^2}{c^2}-i\mu_0\sigma_0\omega)^2-\frac{\mu_0^2\kappa^2\omega^2}{c^2}\Delta p_z^2=0.
\end{align}

\begin{figure*}[!tbp]
\centering
{{\includegraphics[trim=0cm 0cm 0cm 0cm, clip=true,width=18cm, angle=0]{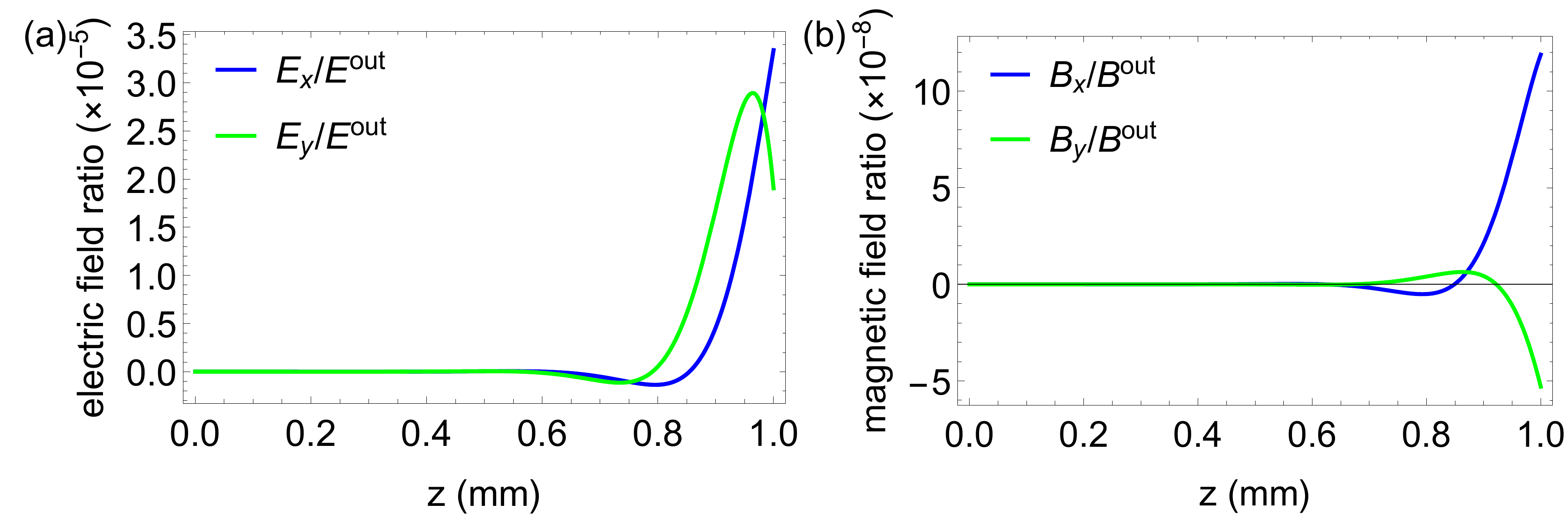}}}
\caption{(a) Electric field components and (b) magnetic field components as a function of distance $z$ inside a semi-infinite slab for $t=0$. Here we set $\Delta\varepsilon=6$ meV, $\Delta p_z c=9.873\cross 10^4$ meV, $\omega=3.0$ GHz and $\sigma_0=10^5$ S/m. The boundary conditions are $E_x(0)=E_y(0)=E^{out}$, $\partial_zE_x(0)=\partial_zE_y(0)=0$ and $B^{out}=\frac{E^{out}}{c}$. The parameters in Eq.~\eqref{eq:main_generalsolution} are $d_1=-26521.3-26519.6i$, $d_2=-7119.09-7121.41i$, $d_3=7124.17+7121.41i$ and $d_4=26516.2+26519.6i$ in units of m$^{-1}$. Here $d_3$ and $d_4$ are both roots with a positive real value, and this gives rise to the exponential growth of the fields with $z$ in this example. We see that the field magnification is strongly enhanced for high conductivity (compare to Fig.~\ref{Fig:1}).}\label{Fig:3}
\end{figure*}

\begin{figure*}[!tbp]
\centering
{{\includegraphics[trim=0cm 0cm 0cm 0cm, clip=true,width=18cm, angle=0]{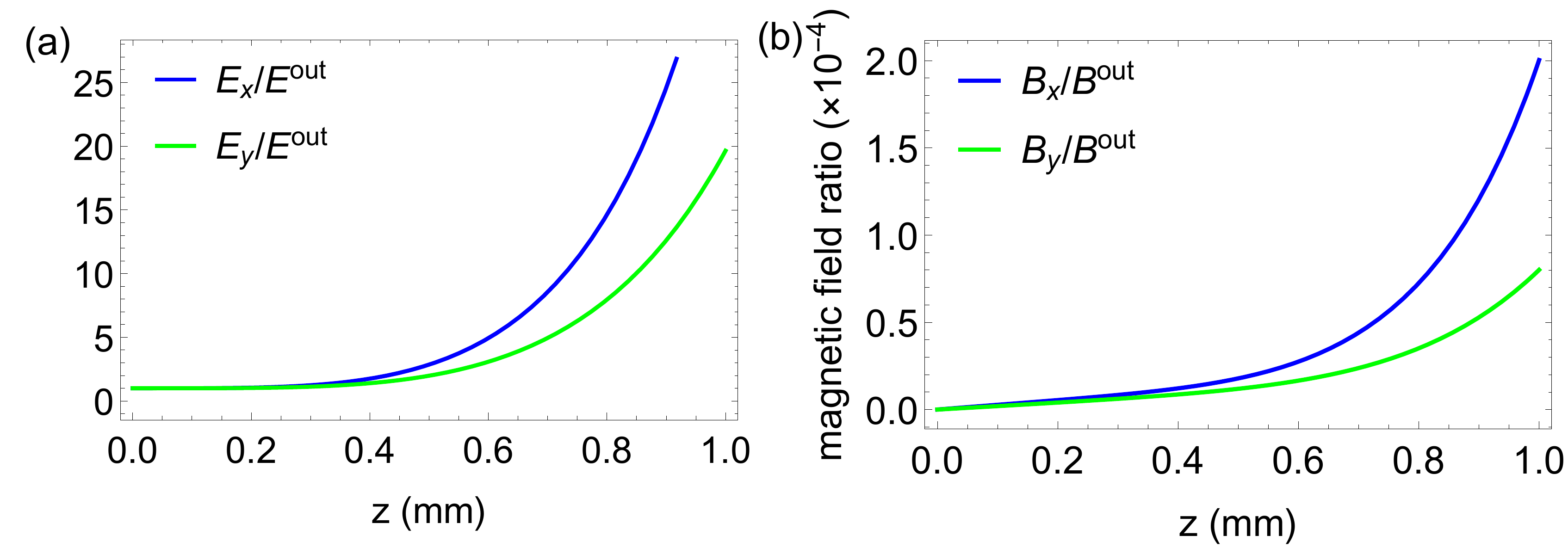}}}
\caption{(a) Electric field components and (b) magnetic field components as a function of distance $z$ inside a semi-infinite slab for $t=0$. Here we set $\Delta\varepsilon=6$ meV, $\Delta p_z c=9.873\cross 10^4$ meV, $\omega=3.0$ GHz and $\sigma_0=10^3$ S/m. The boundary conditions are $E_x(0)=E_y(0)=E^{out}$, $\partial_zE_x(0)=\partial_zE_y(0)=0$ and $B^{out}=\frac{E^{out}}{c}$. The parameters in Eq.~\eqref{eq:main_generalsolution} are $d_1=-4797.92-859.31i$, $d_2=-785.564-4727.14i$, $d_3=785.543+4874.93i$ and $d_4=4797.94+711.514i$ in units of m$^{-1}$. Here $d_3$ and $d_4$ are both roots with a positive real value, and this gives rise to the exponential growth of the fields with $z$ in this example. The magnification of the fields is comparable to that evident in Fig.~\hyperref[Fig:1]{\ref*{Fig:1}}, indicating that for this lower value of the conductivity, the Ohmic term does not contribute significantly to the magnification effect.}\label{Fig:4}
\end{figure*}

Here we make the same parameter choices as in the previous subsection: $\Delta\varepsilon=6$ meV, $\Delta p_z c=\hbar \Delta k_z c=9.873\cross 10^4$ meV, 
and we set the frequency to $\omega=3.0$ GHz. We take the fields outside the WSM ($z<0$) to be
\begin{align}
E_x=E_y&=E^{out}\cos{\frac{\omega}{c}z},\label{eq:main_TDS_vac_E1}\\
B_x=-B_y&=\frac{i}{c}E^{out}\sin{\frac{\omega}{c}z}\label{eq:main_TDS_vac_B2}.
\end{align} 
This choice then implies the following boundary conditions for the fields inside the slab: $E_x(0)=E_y(0)=E^{out}$ and $\partial_zE_x(0)=\partial_zE_y(0)=0$. The typical conductivity of a WSM is smaller than that of a metal. For concreteness, we set $\sigma_0=10^5$ S/m, corresponding to the bulk conductivity of the WSM NbAs~\cite{zhang2019ultrahigh}. We also consider a conductivity that is two orders of magnitude smaller, $10^3$ S/m, to better understand how the conductivity impacts the behavior of the electromagnetic fields.  The resulting electric and magnetic fields inside the WSM for these parameters at $t=0$ are shown in Figs.~\hyperref[Fig:3]{\ref*{Fig:3}} and \hyperref[Fig:4]{\ref*{Fig:4}} with $\sigma_0=10^5$ S/m and $\sigma_0=10^3$ S/m, respectively. In both figures, we choose the same parameters as in the $\vec{j}=0$ case considered in the previous subsection, and we keep the outside fields the same as well.

In Fig.~\hyperref[Fig:3]{\ref*{Fig:3}}, the electric and magnetic fields are both enhanced much more than in the case without the Ohmic current. We find that there are two solutions to Eq.~\eqref{eq_eqn_for_d} that have a positive real part: $d_3=7124.17+7121.41i$ and $d_4=26516.2+26519.6i$ in units of m$^{-1}$. The real part of $d_4$ is much larger than before (see the caption of Fig.~\hyperref[Fig:1]{\ref*{Fig:1}}), and it dominates the growth of the fields. This means that in a real WSM system with a large enough Ohmic conductivity, the magnification of the electromagnetic fields should be more easily detected. Again, we expect that this effect will be weakened in a real sample due to scattering or other effects not accounted for here.

In Fig.~\hyperref[Fig:4]{\ref*{Fig:4}}, the electric and magnetic fields are of a similar magnitude compared to the case of Fig.~\hyperref[Fig:1]{\ref*{Fig:1}}. This means that for low conductivity, the main contribution to the magnification of the fields comes from the non-Ohmic terms. As one can see from the solutions in Fig.~\hyperref[Fig:4]{\ref*{Fig:4}}, although $d_3=785.543+4874.93i$ and $d_4=4797.94+711.514i$ (in units of m$^{-1}$) both have positive real parts, the main contribution is from $d_4$, which is of similar magnitude as in Fig.~\hyperref[Fig:1]{\ref*{Fig:1}}. Therefore, it is legitimate to neglect the Ohmic term and set $\vec{j}=0$.

Before concluding this section, we comment on possible methods to experimentally detect the field magnification effect. To this end, it may be advisable to reach beyond magnetotransport and quantum transport measurements. Instead, it may be more suitable to consider measurements of the magnetic permeability and electrical permittivity for verification of the effects described above. The magnetic permeability quantifies the magnetic field inside the material upon application of an external magnetic field, and experiments can be performed in various sample sizes and applied field configurations. Similarly, the electrical permittivity quantifies the electric field inside the material upon application of an external electric field, and measurements can likewise be performed for various configurations and sample sizes.  The effect of an applied external magnetic field on the electric field inside the material, known as the magnetodielectric effect, and the electric field-induced magnetic permeability are both studied in magnetoelectric materials, and should be considered.  Given the importance of boundary conditions and sample geometry, the measured permeability, permittivity, magnetodielectric coefficient, and electric field-induced magnetic permeability have to be considered as tensors.  Further, the frequency dependence of the tensors can be studied following approaches similar to dielectric spectroscopy.  The frequency-dependence should include detection of higher harmonics to ascertain the possible existence of nonlinear behavior.

\section{Non-dynamical axions and nonlinear chiral magnetic term}\label{sec:ND axion: TID}
In this section we keep the axion non-dynamical, but we consider a different, nonlinear form of the chiral magnetic term. This form was derived from kinetic theory in Ref.~\cite{son2013chiral}. In this approach, one starts from a Boltzmann equation that includes contributions due to a nonzero Berry curvature. These contributions give rise to a chiral magnetic term and an anomalous Hall term. This is the case even for a finite but small chemical potential, such that the system is in a Weyl metal rather than semimetal phase. In this approach, the chiral chemical potential is now proportional to the inner product of the electric and magnetic fields, $\vec{E}\cdot\vec{B}$, while the anomalous Hall term is the same as in Eq.~\eqref{eq: derivation B2}. As we discussed in the previous section, the Ohmic term can be ignored in the limit of low conductivity. Here, we assume this is the case and set $\vec{j}=0$ throughout this section. We consider the case of a nonzero source current in Appendix~\ref{app:NDTI}. The conclusions of this section are largely unaffected by the Ohmic term. The modified Maxwell's equations are then
\begin{align}
\vec{\nabla}\cdot\vec{E}&=-\mu_0c\kappa\Delta\vec{p}\cdot\vec{B},\label{eq:main_TIND_E1}\\
\vec{\nabla}\cross\vec{E}&=0,\label{eq:main_TIND_E2}\\
\vec{\nabla}\cdot\vec{B}&=0,\label{eq:main_TIND_B1}\\
\vec{\nabla}\cross\vec{B}&=\mu_0\sigma_a(\vec{E}\cdot\vec{B})\vec{B}+\frac{\mu_0\kappa}{c}\Delta\vec{p}\cross\vec{E},\label{eq:main_TIND_B2}
\end{align}
where $\sigma_a$ is a constant, and we have again set the source charges and currents to zero: $\rho=0=\vec j$. We see that now the chiral magnetic term in Eq.~\eqref{eq:main_TIND_B2} is nonlinear in $\vec E$ and $\vec B$. Unlike the linear chiral magnetic term in Eq.~\eqref{eq: derivation B2}, a chiral magnetic current is expected to arise even for stationary electric and magnetic fields in this case. In Eqs.~\eqref{eq:main_TIND_E1}-\eqref{eq:main_TIND_B2}, we have already assumed that the fields are time-independent, since this is the case we focus on here. Here, we again assume a single Weyl node pair, although a similar analysis applies for multiple pairs, in which case the anomaly-induced terms in Eqs.~\eqref{eq:main_TIND_E1} and \eqref{eq:main_TIND_B2} receive contributions from each pair. These contributions add linearly~\cite{son2013chiral}, and so effectively this amounts to a simple modification of the coefficients multiplying the electromagnetic fields in these equations. We examine three different geometries: a semi-infinite slab as in the previous section, a case in which the WSM occupies all of space, and a case in which the WSM is an infinite cylindrical wire. In each case, we find that self-consistent solutions do not exist for arbitrary choices of the applied external fields, although solutions can be found in special cases.

\subsection{Semi-infinite slab}

We first consider a semi-infinite slab of WSM occupying $z\geq0$ and where $z<0$ is vacuum. If we consider the case in which the fields outside the slab ($z<0$) are in the $xy$ plane, $\vec{E}=E_x^{out}\hat{x}+E_y^{out}\hat{y}$ and $\vec{B}=B_x^{out}\hat{x}+B_y^{out}\hat{y}$, where $E_x^{out}$ and $E_y^{out}$ are constants, then we immediately run into a problem. From Eq.~\eqref{eq:main_TIND_B2} we see that the number of equations is greater than the number of variables, which leads to a constraint on the fields outside the WSM (see Appendix~\ref{app:NDTI_slab} for details):
\begin{align}
\Delta p_xE_y^{out}=&\Delta p_yE_x^{out}\label{eq:main_TIND_slab_inconsis}.
\end{align}
This imposes a strong constraint on the angle between the electric field outside the WSM and the orientation of the WSM crystal lattice, since the latter determines the orientation of the momentum separation, $\Delta\vec p$, between Weyl nodes. Once we choose the directions of the outside fields, Eq.~\eqref{eq:main_TIND_slab_inconsis} either forces $\Delta\vec p$ to point in a particular direction in the $xy$ plane, or the electric field in the $xy$ plane is forced to be zero. There thus appears to be a fundamental inconsistency in this version of axion electrodynamics, at least as it applies to the semi-infinite slab geometry.

\begin{figure}[!tbp]
\centering
{{\includegraphics[trim=0cm 0cm 0cm 0cm, clip=true,width=8.7cm, angle=0]{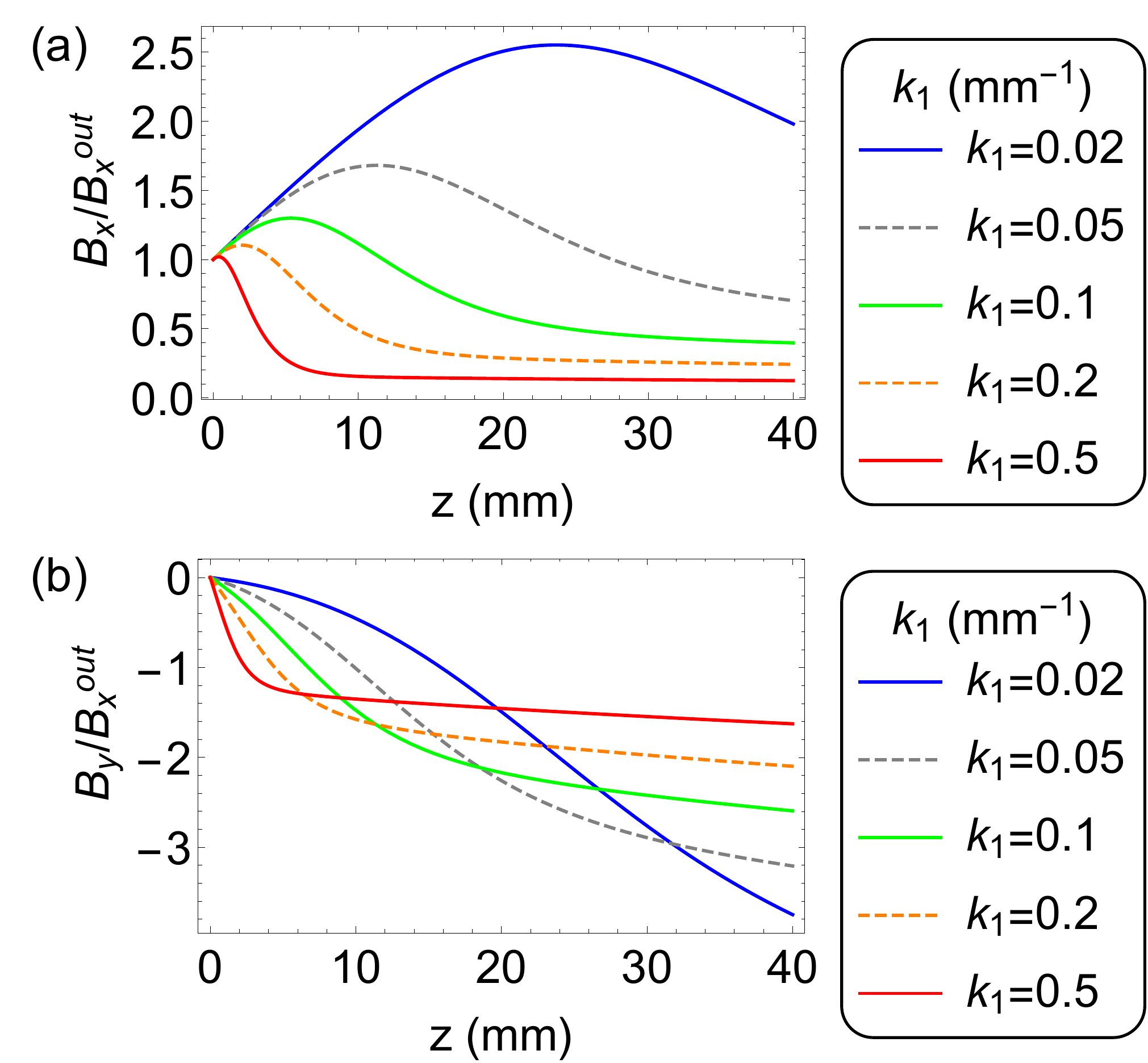}}}
\caption{(a) $x$ and (b) $y$ components of the magnetic field as a function of depth $z$ inside a semi-infinite slab. Here $k_1=\mu_0\sigma_aB_x^{out}E_x^{out}$, $k_2=\frac{\mu_0\kappa\Delta p_zE_x^{out}}{cB_x^{out}}=0.1$ mm$^{-1}$, where $E_x^{out}$ and $B_x^{out}$ are the nonzero components of the applied fields outside the slab.}\label{Fig:5}
\end{figure}

Let us leave this inconsistency aside for the moment and assume that $\Delta p_x=\Delta p_y=0$, in which case the issue is avoided. We then obtain the following equations for the fields inside the WSM:
\begin{align}
E_x&=\text{const.}=E_x^{out},\qquad
E_y=\text{const.}=E_y^{out},\\
B_z&=0,\qquad
\partial_zE_z=0,\\
\partial_zB_y&=-\mu_0\sigma_a(E_x^{out}B_x+E_y^{out}B_y)B_x+\frac{\mu_0\kappa}{c}\Delta p_z E_y^{out},\\
\partial_zB_x&=\mu_0\sigma_a(E_x^{out}B_x+E_y^{out}B_y)B_y+\frac{\mu_0\kappa}{c}\Delta p_zE_x^{out}.
\end{align}
Here, we have used that the fields are continuous across the surface, which is shown in Appendix~\ref{app:NDTI_slab}. 
Since we are assuming there is no $E_z$ component outside of the sample, we have
$E_z=E_z^{out}=0$.
Suppose that we also have $E_y^{out}=0=B_y^{out}$, i.e., the applied electric and magnetic fields are parallel and lie in the $x$ direction, transverse to the surface. The last two equations above then become
\begin{align}
\partial_zB_y&=-\mu_0\sigma_aE_x^{out}B_x^2,\\
\partial_zB_x&=\mu_0\sigma_aE_x^{out}B_xB_y+\frac{\mu_0\kappa}{c}\Delta p_zE_x^{out}.
\end{align}
We can render these equations dimensionless by dividing both sides by $B_x^{out}$ and then defining $k_1=\mu_0\sigma_aB_x^{out}E_x^{out}$ and $k_2=\frac{\mu_0\kappa\Delta p_zE_x^{out}}{cB_x^{out}}$. Because the fields must be continuous at the boundary, we impose $B_{x}(0)=B_x^{out}$ and $B_{y}(0)=0$. We show the solution of these equations in Fig.~\hyperref[Fig:5]{\ref*{Fig:5}}. As one can see, although we have set $B_y(0)=0$ at the surface, the equations still yield a nonzero $B_y$ inside the WSM. In addition, the magnetic field component in the $x$ direction decreases with increasing depth into the slab. In the limit of very large $z$, $B_x$ becomes arbitrarily close to zero. These solutions reveal that the electric and magnetic fields are trying to become perpendicular at large $z$. Thus, the fields inside the slab arrange themselves in such a way that the CME is suppressed. Similar results were found in Ref.~\cite{barnes2016electromagnetic} in the absence of the AHE term.

Let us now consider the case where the outside fields are in the $z$ direction, i.e., $\vec{E}=E_z^{out}\hat{z}$ and $\vec{B}=B_z^{out}\hat{z}$. We again assume $\Delta p_x=\Delta p_y=0$. As shown in Appendix~\ref{app:NDTI_slab}, solutions only exist if
$B_z=B_z^{out}=0$.
This is a contradiction, since we assumed $B_z^{out}\neq 0$ at the outset, and we should be free to choose the applied fields any way we like. This again suggests that there may be an intrinsic inconsistency with Eqs.~\eqref{eq:main_TIND_E1}-\eqref{eq:main_TIND_B2}. Next, we examine whether similar issues arise for other geometries.

\subsection{Whole space case}
Now we consider the case where the whole space is a WSM. In this case, all the fields must be constant due to symmetry. Eqs.~\eqref{eq:main_TIND_E1}-\eqref{eq:main_TIND_B2} reduce to
\begin{align}
\vec{\nabla}\cdot\vec{E}&=-\mu_0c\kappa\Delta\vec{p}\cdot\vec{B}=0,\\
\vec{\nabla}\cross\vec{B}&=\mu_0\sigma_a(\vec{E}\cdot\vec{B})\vec{B}+\frac{\mu_0\kappa}{c}\Delta\vec{p}\cross\vec{E}=0.
\end{align}
If we pick the direction of the Weyl node momentum separation to be $\hat{z}$, $\Delta\vec{p}=\Delta p_z \hat{z}$, we have $\vec{B}=(B_x,B_y,0)$ from the first equation. After some steps shown in Appendix~\ref{app:NDTI_whole}, we obtain
\begin{equation}
E_x=E_y=0,\qquad
E_z=\text{const.}
\end{equation}
Thus the conclusion for this case is that the electric field must be parallel to $\Delta\vec{p}$, and the magnetic field is perpendicular to it. Therefore, the CME disappears automatically in this case.

\subsection{Infinite cylindrical wire}
Next, we study an infinite cylindrical wire with radius $R$. We use cylindrical coordinates, taking the axis of the wire to lie in the $z$ direction and defining $r$ to be the radial coordinate. The wire is a WSM, and outside is vacuum. For simplicity, we choose the Weyl node separation in momentum space to be in the $z$ direction: $\Delta p_x=\Delta p_y=0$ and $\Delta p_z\ne0$. With these assumptions and switching to cylindrical coordinates, Eqs.~\eqref{eq:main_TIND_E1}-\eqref{eq:main_TIND_B2} become (see Appendix~\ref{App_B3} for details):
\begin{align}
E_z=E_z^{out},\quad
E_\phi=0,\quad
B_r&=0,\\
\frac{1}{r}\frac{\partial}{\partial r}(rE_r)+\mu_0c\kappa\Delta p_zB_z&=0,\label{eq:main_TIND_cylinder_E3}\\
\frac{\partial B_z}{\partial r}+\mu_0\sigma_aE_z^{out}B_zB_\phi+\frac{\mu_0\kappa}{c}\Delta p_zE_r&=0,\\
-\frac{1}{r}\frac{\partial}{\partial r}(rB_\phi)+\mu_0\sigma_aE_z^{out}B_z^2&=0.\label{eq:main_TIND_cylinder_B3}
\end{align}
A similar set of equations was solved in Ref.~\cite{barnes2016electromagnetic}, although there the AHE term was neglected. We first revisit this case before solving the full equations with the AHE term present, as we will find that both cases exhibit common pathologies. The solution that was obtained in Ref.~\cite{barnes2016electromagnetic} has a diverging electric field along the axis of the wire, $E_r\to\infty$ as $r\to0$, as we now show. Inside the WSM, $B_z$ was found to be
\begin{align}
B_z=\frac{2B_0\Lambda k}{r^2+k^2},
\end{align}
where $\Lambda=(\mu_0\sigma_a E_{0}B_0)^{-1}$ and $k=\Lambda+\sqrt{\Lambda^2-R^2}$, with applied fields $\vec{E}^{out}=E_0\hat z$, $\vec{B}^{out}=B_0\hat z$ outside the wire.
Plugging this result for $B_z$ into Eq.~\eqref{eq:main_TIND_cylinder_E3}, one obtains
\begin{align}
\frac{\partial}{\partial r}(rE_r)&=-\frac{2\mu_0c\kappa\Delta p_zB_0\Lambda k r}{r^2+k^2}\\
\Rightarrow E_r&=-\frac{2\mu_0c\kappa\Delta p_zB_0\Lambda k}{r}\int{\frac{rdr}{r^2+k^2}}\nonumber\\
&=-\frac{\mu_0c\kappa\Delta p_zB_0\Lambda k}{r}[\ln(r^2+k^2)+C_1],
\end{align}
which is singular at $r=0$. 

\begin{figure}[!tbp]
\centering
{{\includegraphics[trim=0cm 0cm 0cm 0cm, clip=true,width=8.7cm, angle=0]{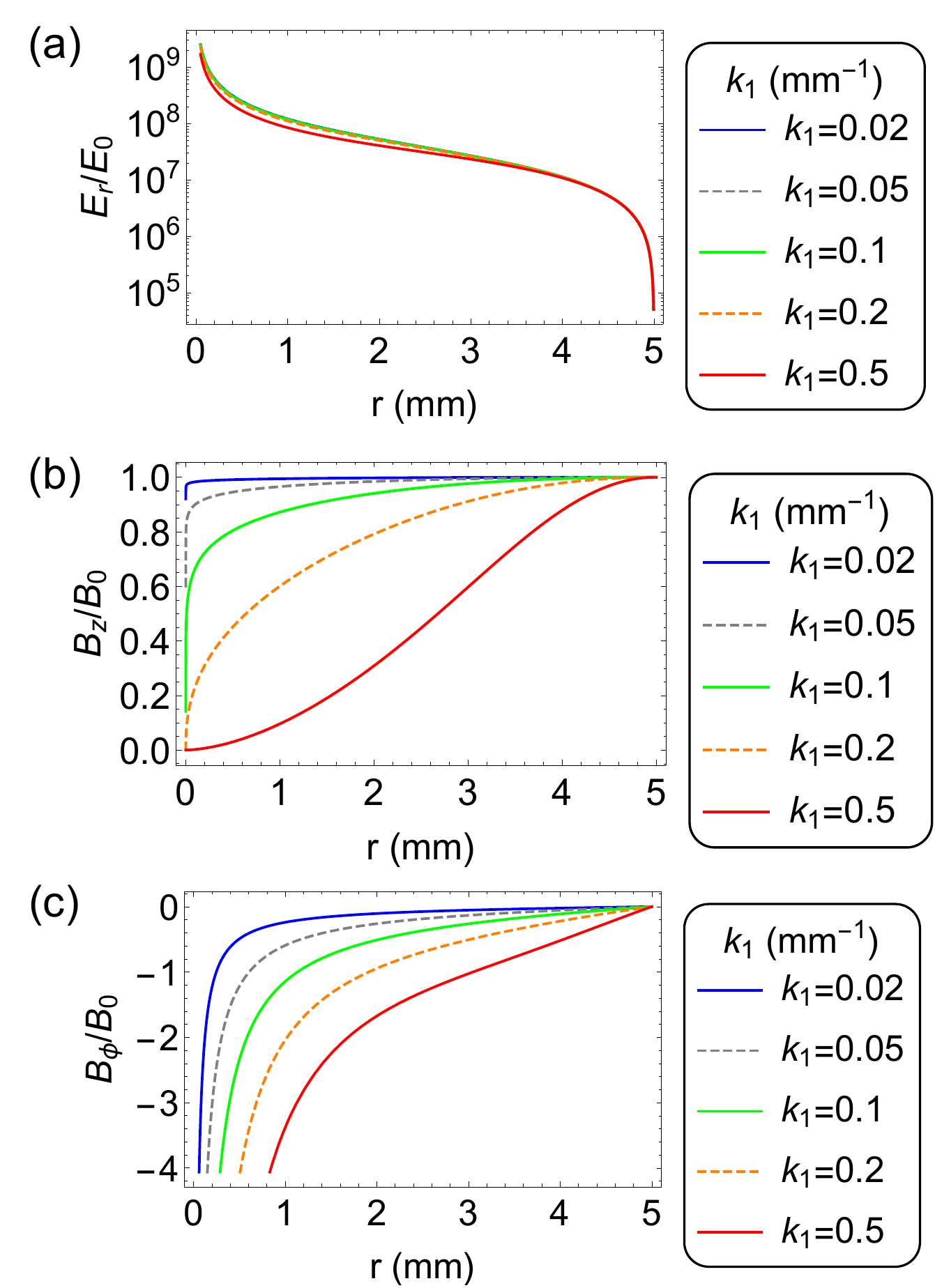}}}
\caption{Three different electric and magnetic field components as a function of the radius $r$ inside an infinite cylindrical WSM wire without the anomalous Hall term [Eqs.~\eqref{eq:main_cylinder_noAHE_1}-\eqref{eq:main_cylinder_noAHE_3}]. Here $E_{\phi}^{in}=E_{\phi}^{out}=0$, $E_z^{in}=E_z^{out}=E_0$, $B_r^{in}=B_r^{out}=0$, $k_1=\mu_0\sigma_a E_{0}B_0$ and $k_3=\frac{\mu_0c\kappa\Delta p_zB_0}{E_0}=10^7$mm$^{-1}$. At the boundary $R=5$ mm, we set $E_r(R)=0.5E_{0}$, $B_z(R)=B_{0}$ and $B_\phi(R)=0$. These solutions show that the electric and magnetic fields become perpendicular to each other at the center of the wire.}\label{Fig:6}
\end{figure}

The singular behavior of the solution above persists for arbitrary choices of the outside fields.  Define $k_1=\mu_0\sigma_a E_{0}B_0$ and $k_3=\frac{\mu_0c\kappa\Delta p_zB_0}{E_0}$, Eqs.~\eqref{eq:main_TIND_cylinder_E3}-\eqref{eq:main_TIND_cylinder_B3} become
\begin{align}
\frac{1}{r}\frac{\partial}{\partial r}(r\frac{E_r}{E_0})+k_3\frac{B_z}{B_0}&=0,\label{eq:main_cylinder_noAHE_1}\\
\frac{\partial}{\partial r}\frac{B_z}{B_0}+k_1\frac{B_z}{B_0}\frac{B_\phi}{B_0}&=0,\label{eq:main_cylinder_noAHE_2}\\
-\frac{1}{r}\frac{\partial}{\partial r}(r\frac{B_\phi}{B_0})+k_1\frac{B_z^2}{B_0^2}&=0.\label{eq:main_cylinder_noAHE_3}
\end{align}
Here, $E_0$ and $B_0$ parameterize the fields outside the wire. We should be able to choose the outside fields as desired. In Appendix~\ref{App_B3}, we show that the fields must be continuous at the surface of the wire, meaning that we should be free to choose the boundary conditions of the fields at $r=R$; these boundary values then determine the fields inside the WSM. As a concrete example, we set $R=5$~mm and choose $E_r(R)=0.5E_{0}$, $B_z(R)=B_{0}$, and $B_\phi(R)=0$, which correspond to a radial electric field and an axial magnetic field outside. The solution is shown in Fig.~\hyperref[Fig:6]{\ref*{Fig:6}}. In these solutions, we do not restrict ourselves to finite values for $B_\phi$ at $r=0$ as in Ref.~\cite{barnes2016electromagnetic}, since the singularity at $r=0$ arises regardless of how $B_\phi$ behaves along the cylinder axis. We can identify two possible explanations for these unavoidable divergences at $r=0$: (i) The axion equations may be intrinsically problematic; (ii) In this cylindrical WSM, the axial anomaly creates an effective line charge and current at $r=0$. We do not currently see a way to establish which interpretation is correct. Interestingly, notice that since $E_z$ is constant inside the wire, and $B_z$ decreases while the magnitude of $B_\phi$ increases as $r\to0$, we again find that the electric and magnetic fields become perpendicular as we go further into the WSM, just as we saw for the semi-infinite slab above.

\begin{figure}[!tbp]
\centering
{{\includegraphics[trim=0cm 0cm 0cm 0cm, clip=true,width=8.7cm, angle=0]{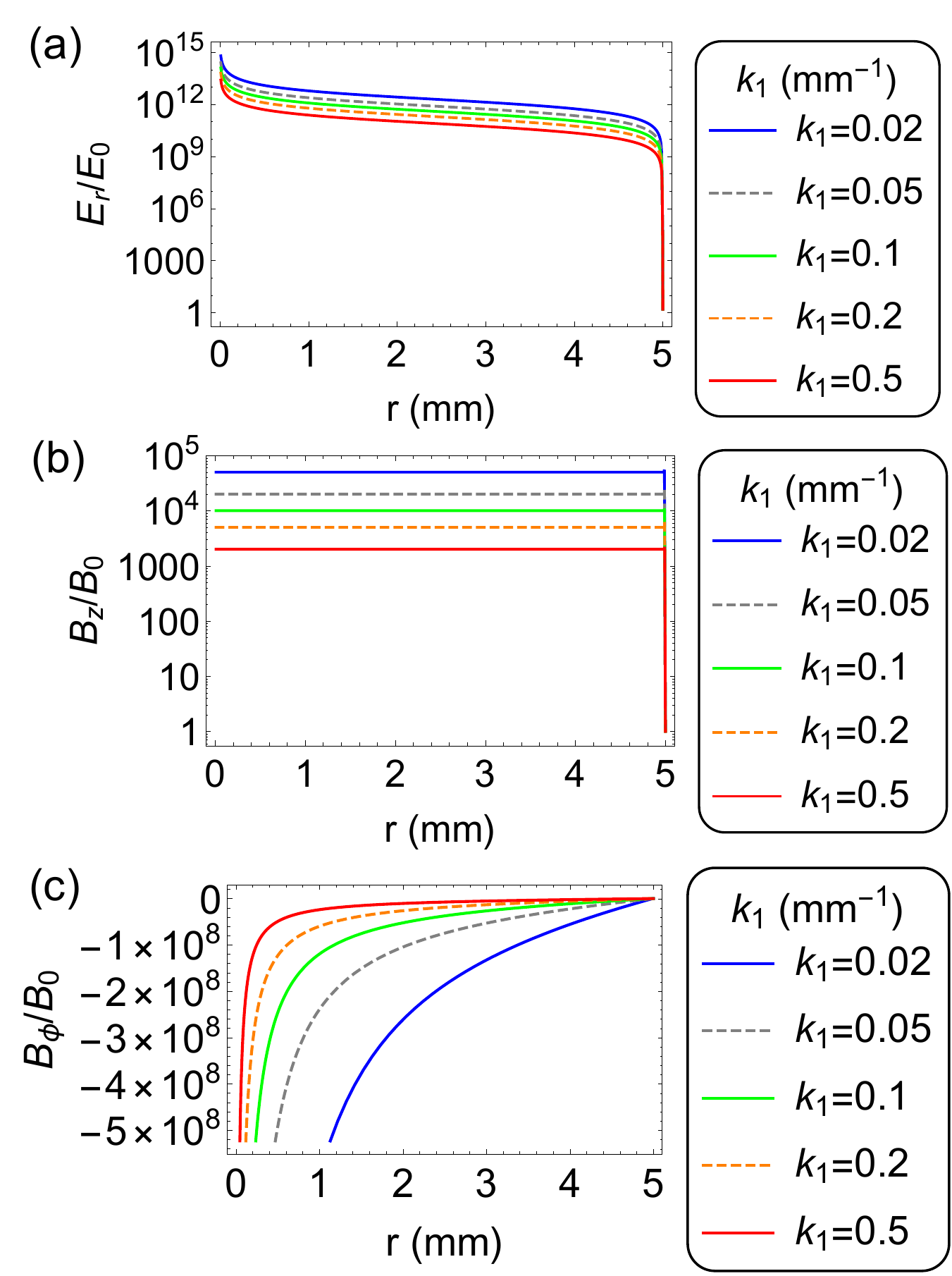}}}
\caption{Three different electric and magnetic field components as a function of the radius $r$ inside an infinite cylindrical WSM wire with the anomalous Hall term restored [Eqs.~\eqref{eq:main_cylinder_withAHE_1}-\eqref{eq:main_cylinder_withAHE_3}]]. Here $E_{\phi}^{in}=E_{\phi}^{out}=0$, $E_z^{in}=E_z^{out}=E_0$, $B_r^{in}=B_r^{out}=0$, $k_1=\mu_0\sigma_a E_{0}B_0$, $k_3=\frac{\mu_0c\kappa\Delta p_zB_0}{E_0}=10^7$ mm$^{-1}$, and $k_2=\frac{\mu_0\kappa\Delta p_zE_0}{cB_0}=0.1$ mm$^{-1}$. At the boundary $R=5$ mm, we set $E_r(R)=0.5E_{0}$, $B_z(R)=B_{0}$ and $B_\phi(R)=0$. In this case, the electric and magnetic fields do not become perpendicular at the center of the wire. Instead, $B_z$ tends to a constant at $r=0$ that depends on the parameters $k_i$.}\label{Fig:7}
\end{figure}

Now we return to the full axion electrodynamics equations with the anomalous Hall term restored. If we define $k_1=\mu_0\sigma_a E_{0}B_0$, $k_3=\frac{\mu_0c\kappa\Delta p_zB_0}{E_0}$, and $k_2=\frac{\mu_0\kappa\Delta p_zE_0}{cB_0}$, Eqs.~\eqref{eq:main_TIND_cylinder_E3}-\eqref{eq:main_TIND_cylinder_B3} become
\begin{align}
\frac{1}{r}\frac{\partial}{\partial r}(r\frac{E_r}{E_0})+k_3\frac{B_z}{B_0}&=0,\label{eq:main_cylinder_withAHE_1}\\
\frac{\partial}{\partial r}\frac{B_z}{B_0}+k_1\frac{B_z}{B_0}\frac{B_\phi}{B_0}+k_2\frac{E_r}{E_0}&=0,\\
-\frac{1}{r}\frac{\partial}{\partial r}(r\frac{B_\phi}{B_0})+k_1\frac{B_z^2}{B_0^2}&=0.\label{eq:main_cylinder_withAHE_3}
\end{align}
All the solutions of these equations face the same problem as before, namely they exhibit singularities at $r=0$. We show one example in Fig.~\hyperref[Fig:7]{\ref*{Fig:7}}. Here we choose $R=5$ mm, $E_r(R)=0.5E_{0}$, $B_z(R)=B_{0}$, and $B_\phi(R)=0$, corresponding to radial electric and magnetic fields outside the wire. We also find that even when turning off $E_r$ outside, this component still increases inside the wire and diverges as $r\to0$. Thus, singularities in the fields along the cylinder axis again appear to be unavoidable. However, unlike the case above where we neglected the AHE term, now the electric and magnetic fields are no longer becoming perpendicular to each other as $r\to0$ in these solutions. Instead, $B_z(0)$ is a nonzero constant that depends on the parameters $k_i$.

In summary, we find that in the cylindrical wire case, we can always find solutions for the electric and magnetic fields inside the wire. This is in contrast to the semi-infinite slab, where we saw that when the chiral magnetic term is nonlinear, self-consistent solutions are not available. However, the fields inside the wire necessarily exhibit singularities along the wire axis.

\section{Dynamical axions}\label{sec:D axion}
In the previous sections, we considered two different versions of axion electrodynamics. Both are based on a non-dynamical axion, i.e., the axion arises as a background field that interacts with the electric and magnetic fields. However, axions in topological materials can have their own dynamics~\cite{taguchi2018electromagnetic,wang2013chiral}. Ref.~\cite{wang2013chiral} showed that dynamical axions can arise in WSMs, for example as fluctuations in the phase of an order parameter associated with a charge density wave. In this section, we consider a third version of axion electrodynamics in which the axion is an independent, dynamical field.

Allowing the axion to be dynamical introduces an additional, fifth equation: the equation of motion for the axion. This equation can be derived from a Lagrangian density as in Eqs.~\eqref{eq:L0} and \eqref{eq:L1}, except that now we introduce an additional kinetic term for the pseudo-scalar axion field $\theta$:
\begin{align}
\mathcal{L}_{\theta_2}=\frac{1}{2}\kappa_0\partial_\alpha\theta\partial^\alpha\theta=\frac{1}{2}\kappa_0\partial_\alpha\theta\partial_\beta\theta\eta^{\alpha\beta},
\end{align}
where $\kappa_0$ is a constant. Combining this with Eqs.~\eqref{eq:L0} and~\eqref{eq:L1}, our total Lagrangian density is
\begin{align}\label{eq:totalLagrangian}
\mathcal{L}=\mathcal{L}_0+\mathcal{L}_{\theta_1}+\mathcal{L}_{\theta_2}.
\end{align}
In addition to Eqs.~\eqref{eq: derivation E1}-\eqref{eq: derivation B2}, the Euler-Lagrange equations now also give the equation of motion for the axion:
\begin{align}
\partial_\nu\partial^\nu\theta=\frac{\kappa}{\kappa_0c}\vec{E}\cdot\vec{B}.\label{eq:Dynamical_kinetic}
\end{align}
For simplicity, we set the source terms to zero in this section:, $\vec{j}=0$, $\rho=0$. Restricting attention to stationary $\vec{E}$ and $\vec{B}$ fields, we have the following version of axion electrodynamics:
\begin{align}
&\vec{\nabla}\cdot\vec{E}=-\mu_0c\kappa\vec{\nabla}\theta\cdot\vec{B},\label{eq:dynamical main_1}\\
&\vec{\nabla}\cross\vec{E}=0,\\
&\vec{\nabla}\cdot\vec{B}=0,\\
&\vec{\nabla}\cross\vec{B}=\frac{\mu_0\kappa}{c}(\partial_t\theta\vec{B}+\vec{\nabla}\theta\cross\vec{E}),\label{eq:dynamical main_4}\\
&\frac{1}{c^2}\partial_t^2\theta-\vec{\nabla}^2\theta=\frac{\kappa}{\kappa_0c}\vec{E}\cdot\vec{B}.\label{eq:dynamical main_5}
\end{align}
Here as in the previous sections, we assume a single Weyl node pair. Multiple pairs would introduce additional axion fields and corresponding kinetic equations of the form of Eq.~\eqref{eq:dynamical main_5}. Because our focus is on the self-consistency of the axion equations, we consider the simplest case of a single node pair to more clearly highlight the issues that arise.
Notice that the above equations do not contain any information about the band structure of the WSM. In Ref.~\cite{wang2013chiral}, the Weyl separations appear only implicitly as a shift of the derivatives of the axion field $\partial_\mu\theta$. We return to this point shortly. Let us first focus on solving the equations above.

As a concrete example, we again consider a semi-infinite slab of WSM occupying the upper half-space $z\ge0$. Because we are focusing on the case where the electric and magnetic fields are stationary, Eqs.~\eqref{eq:dynamical main_4} and \eqref{eq:dynamical main_5} imply that $\theta$ is at most a linear function of $t$. Futhermore, for the semi-infinite slab symmetry, $\vec{\nabla}\theta$ can depend on $z$ only. Therefore, the most general form of $\theta$ is
\begin{align}\label{eq:theta_ansatz}
\theta&=f_x(z) x+f_y(z) y+f_{t,0} t+\widetilde{\theta}(z),
\end{align}
where $f_{t,0}$ is a constant due to the fact that $\partial_t\theta$ does not depend on $t$. The symmetry of the slab geometry also implies that $\partial_z\theta$ depends on $z$ only. This in turn means that $f_x(z)=f_{x,0}$ and $f_y(z)=f_{y,0}$ are constants. Denoting $\partial_z\widetilde{\theta}(z)=f_z(z)$, the derivatives of the axion thus have the following generic form for the semi-infinite slab geometry in the case of stationary electric and magnetic fields:
\begin{equation}
\vec{\nabla}\theta=\vec{f}=f_{x,0}\hat x+f_{y,0}\hat y+f_z(z)\hat z,\quad
\partial_t\theta=f_{t,0}.
\end{equation}
Eqs.~\eqref{eq:dynamical main_1}-\eqref{eq:dynamical main_5} then reduce to the following set of algebraic and ordinary differential equations:
\begin{align}
E_x=&E_x^{out},\quad\label{eq:case7_const_E_x}
E_y=E_y^{out},\quad
B_z=B_z^{out},\\
0=&f_{t,0}B_z^{out}+f_{x,0}E_y^{out}-f_{y,0}E_x^{out},\label{eq:case7_constrain_ft}\\
\partial_zE_z=&-\mu_0c\kappa(f_{x,0}B_x+f_{y,0}B_y+f_zB_z^{out}),\label{eq:case7_Ez}\\
\partial_zB_x=&\frac{\mu_0\kappa}{c}(f_{t,0}B_y+f_zE_x^{out}-f_{x,0}E_z),\\
\partial_zB_y=&-\frac{\mu_0\kappa}{c}(f_{t,0}B_x+f_{y,0}E_z-f_zE_y^{out}),\\
\partial_zf_z=&-\frac{\kappa}{\kappa_0c}(E_x^{out}B_x+E_y^{out}B_y+E_zB_z^{out})\label{eq:case7_fz}.
\end{align}
Here, we have used that all components of the electric and magnetic fields are again continuous across the surface, as follows from arguments similar to those used in the context of the other two versions of axion electrodynamics considered in this work.
Eq.~\eqref{eq:case7_constrain_ft} gives a constraint for the axion derivative $f_\mu$; the effect of this constraint depends on how we choose the applied fields outside the WSM, as is evident in the examples given below. The examples we consider include the case where the applied fields are orthogonal to the WSM surface (Sec.~\ref{case A}), and where they are parallel to the surface (Sec.~\ref{case B}). We also examine energy conservation in Sec.~\ref{energy cons}, where we find evidence that time-independent solutions should not exist in the case of a dynamical axion.

\subsection{$\vec{E}, \vec{B} \parallel \hat{z}$ outside of the WSM}\label{case A}
When both the electric and magnetic fields are orthogonal to the surface, we have
$
E_x^{out}=E_y^{out}=0,
$
and so Eq.~\eqref{eq:case7_constrain_ft} implies that the CME term vanishes, $f_{t,0}=0$, when $B_z^{out}\ne0$.
The other boundary conditions are $B_x(0)=B_y(0)=0$ and $E_z(0)=E_z^{out}$. The solutions to Eqs.~\eqref{eq:case7_Ez}-\eqref{eq:case7_fz} in this case are (see Appendix~\ref{app: case A})
\begin{align}
B_x=&\frac{\kappa^2\mu_0^2f_{x,0}f_{z,0}B_z^{out}}{D^2}(-1+\cosh{D z})-\frac{E_z^{out}}{c D}\sinh{D z},\\
B_y=&\frac{\kappa^2\mu_0^2f_{y,0}f_{z,0}B_z^{out}}{D^2}(-1+\cosh{D z})-\frac{E_z^{out}}{c D}\sinh{D z},\\
E_z=&E_z^{out}\cosh{D z}-\frac{B_z^{out}c f_{z,0}\kappa\mu_0}{D}\sinh{D z},\\
f_z=&\frac{f_{z,0}(\kappa_0 D^2-{B_z^{out}}^2\kappa^2\mu_0+{B_z^{out}}^2\kappa^2\mu_0\cosh{D z})}{\kappa_0 D^2}\nonumber\\
&-\frac{B_z^{out}E_z^{out}\kappa\sinh{D z}}{c\kappa_0 D},
\end{align}
where we have defined
\begin{align}
D^2=\frac{\kappa^2\mu_0[{B_z^{out}}^2+(f_{x,0}^2+f_{y,0}^2)\kappa_0\mu_0]}{\kappa_0}.
\end{align}
Here, we allow for the possibility of a finite jump in the derivative of the axion at the surface: $f_z(0)=f_{z,0}$. We see that the fields grow exponentially with $z$, where the rate of growth is set by $D$, which depends on the applied magnetic field and on the transverse derivatives of the axion. This growth should ultimately saturate for a finite slab. Aside from this unbounded growth, which is a simple consequence of the infinite slab geometry considered here, no pathologies appear to arise in this case. 

Following Ref.~\cite{wang2013chiral}, one would expect the scale of spatial and temporal variations in $\theta$ to depend on the Weyl momentum and energy separations $\Delta\vec{p}$ and $\Delta\varepsilon$. Since the space-time dependence of $\theta$ is determined by the boundary values $f_{x,0}$, $f_{y,0}$, $f_{z,0}$, and $f_{t,0}$, it follows that these quantities should depend on the Weyl momentum and energy separations, and thus they depend on the type of WSM under consideration. It is not clear whether the precise relationship between the boundary values of $f_\mu$ and the Weyl node separation can be obtained in closed form.

\subsection{$\vec{E}, \vec{B} \parallel \hat{x}$ outside of the WSM}\label{case B}
Now we consider the case where the fields outside the slab are parallel to the WSM surface. In particular, we will take them to both point in the $x$ direction for concreteness. Explicitly, we have
$B_z^{out}=E_y^{out}=0$,
and from the constraint in Eq.~\eqref{eq:case7_constrain_ft}, we can see that $f_{t,0}$ and $f_{x,0}$ are no longer restricted, while
$f_{y,0}=0$.
The remaining boundary conditions in this case are $B_x(0)=B_x^{out}$, $B_y(0)=0$, $E_z(0)=0$, and we again allow for a possible discontinuity in $f_z(z)$ at the surface: $f_z(0)=f_{z,0}$. The solutions to Eqs.~\eqref{eq:case7_Ez}-\eqref{eq:case7_fz} in this case are (see Appendix~\ref{app: case B})
\begin{align}
B_x=&B_x^{out}\cos{D_0z}+\frac{E_x^{out}f_{z,0}\kappa\mu_0}{c D_0}\sin{D_0 z},\\
B_y=&\frac{f_{t,0}E_x^{out}f_{z,0}\kappa^2\mu_0^2}{c^2D_0^2}(-1+\cos{D_0 z})\nonumber\\
&-\frac{B_x^{out}f_{t,0}\kappa\mu_0}{c D_0}\sin{D_0 z},\\
E_z=&\frac{f_{x,0}E_x^{out}f_{z,0}\kappa^2\mu_0^2}{D_0^2}(-1+\cos{D_0 z})\nonumber\\
&-\frac{B_x^{out}c f_{x,0}\kappa\mu_0}{D_0}\sin{D_0 z},\\
f_z=&\frac{\kappa}{c^2D_0^2\kappa_0}[c D_0f_{z,0}\sqrt{\kappa_0\mu_0}-{E_x^{out}}^2f_{z,0}\kappa\mu_0(1-\cos{D_0 z})\nonumber\\
&-B_x^{out}E_x^{out}c D_0\sin{D_0 z}],
\end{align}
where now
\begin{align}
D^2=-D_0^2=\frac{-\kappa^2\mu_0[{E_x^{out}}^2+\kappa_0\mu_0(f_{t,0}^2-c^2f_{x,0}^2)]}{c^2\kappa_0}.
\end{align}
The solutions in this case exhibit oscillating behavior for all choices of the remaining parameters. Again, no inconsistencies appear in this case.

\subsection{Energy conservation}\label{energy cons}
Now let us check whether energy is conserved in a WSM described by a dynamical axion field subject to stationary electric and magnetic fields. In dielectric media, the energy-momentum tensor of the electromagnetic fields might not be conserved. This is related to the long-standing Abraham–Minkowski controversy, which continues to be debated~\cite{kemp2011resolution,mansuripur2010resolution,wang2011crucial}. While the electromagnetic stress-energy tensor is generally not conserved in the presence of matter, here we still expect it to be conserved because the material has been replaced by an axion field, and so we are effectively dealing with axion electrodynamics in vacuum. We can obtain the stress-energy tensor from the Lagrangian density $\mathcal{L}$ in Eq.~\eqref{eq:totalLagrangian}~\cite{carroll2019spacetime}:
\begin{align}
T^{\mu\nu}=\frac{\partial\mathcal{L}}{\partial(\partial_\mu A_{\sigma})}\partial^{\nu}A_{\sigma}+\frac{\partial\mathcal{L}}{\partial(\partial_\mu\theta)}\partial^{\nu}\theta-\eta^{\mu\nu}\mathcal{L},
\end{align}
or more explicitly,
\begin{align}
T^{\mu\nu}=&\kappa_0\partial^\mu \theta\partial^\nu \theta-\frac{1}{\mu_0}F^{\mu\gamma}\partial^\nu A_\gamma\nonumber\\
&-\frac{\kappa}{2}\theta\varepsilon^{\mu\gamma\sigma\lambda}F_{\sigma\lambda}\partial^\nu A_\gamma-\eta^{\mu\nu}\mathcal{L}.
\end{align}
After simplification, the energy density $T^{00}$ reads
\begin{align}
T^{00}=\frac{1}{2}(\varepsilon_0\vec{E}^2+\frac{1}{\mu_0}\vec{B}^2)+\frac{1}{2}\kappa_0(\partial_0\theta\partial_0\theta+\partial_i\theta\partial_i\theta),
\end{align}
where there is an implicit sum over the index $i$. The first term is the energy density of the electromagnetic field, while the second term is the energy density of the dynamical axion field. On one hand, if one does not have a kinetic term in the Lagrangian, one would only get the energy density of the electromagnetic fields, which is the case considered in Sec.~\ref{sec:ND axion: TD}. The energy density is continuous across the boundary in this case since the electromagnetic fields are continuous. On the other hand, when one includes the kinetic terms for $\theta$, demanding that the energy density be continuous across the boundary requires the kinetic term to vanish at the boundary:
\begin{align}
f_{x,0}^2+f_{y,0}^2+f_{z,0}^2+\frac{1}{c^2}f_{t,0}^2=0.
\end{align}
This can only be satisfied if all the axion derivatives vanish at the surface:
\begin{align}
f_{x,0}=f_{y,0}=f_{z,0}=f_{t,0}=0.
\end{align}
Referring back to Eqs.~\eqref{eq:dynamical main_4} and \eqref{eq:theta_ansatz}, we see that this forces the chiral magnetic term to vanish. We also see that the constraint shown in Eq.~\eqref{eq:case7_constrain_ft} holds automatically and does not place any restriction on the electromagnetic fields. Although $f_{x,0}$ and $f_{y,0}$ will always be zero inside of the WSM, $f_z$ could still be nonzero. Therefore, nontrivial solutions can still be obtained. However, these solutions only provide signatures of the AHE term. Perhaps one way to obtain a response from the chiral magnetic term would be to relax the assumption of static applied fields and to instead consider time-dependent fields. Where or not self-consistent solutions can be obtained in this case will be investigated in future work.

Before we finish this section, it is worth considering whether the solutions to the dynamical axion equations have any relation to the solutions obtained in Sec.~\ref{sec:ND axion: TD} in the case of a non-dynamical axion (with a linear chiral magnetic term). Naively, one can try to insert the latter into the dynamical axion equations. However, one immediately finds that this does not work, because the left-hand side of Eq.~\eqref{eq:dynamical main_5} evaluates to zero, yielding a constraint on the electric and magnetic fields (they must be orthogonal), while the other equations remain the same. Even if we chose the applied fields to be orthogonal to each other, it is not guaranteed that they will remain orthogonal inside the WSM. Indeed, we have checked whether $\vec{E}\cdot\vec{B}=0$ is approximately obeyed by the solutions of Sec.~\ref{sec:ND axion: TD}, and we found that $\vec{E}\cdot\vec{B}$ instead grows quickly with depth into the WSM. (Note that this is unlike the solutions obtained in the case of a nonlinear chiral magnetic term, where in Sec.~\ref{sec:ND axion: TID} we found several instances in which $\vec{E}\cdot\vec{B}\to0$ as $z\to\infty$.) Therefore, there does not appear to be a sense in which the non-dynamical axion solutions (Sec.~\ref{sec:ND axion: TD}) approximate the dynamical axion solutions obtained in the present section.

\section{Conclusions}\label{sec:Conclu}
Whether or not the axial anomaly exists in WSMs remains a subtle question. The motivation for our work is to identify alternative diagnostics based on electromagnetic signatures that could be exploited to experimentally confirm the presence of an anomaly. To this end, we considered three versions of axion electrodynamics that have been put forward in the literature. In each case, we attempted to solve the equations in simple geometries.

In the first version, we started from an effective action for non-dynamical axions given by Refs.~\cite{zyuzin2012topological,chen2013axion}. In the case of a semi-infinite slab, we found that the magnetic field inside the WSM can be magnified substantially assuming the Weyl node momentum separation and the frequency of the applied fields are both sufficiently large, which happens with or without the Ohmic current term. We also found that when the conductivity is sufficiently large, this magnification effect is further enhanced. This potentially provides a detectable signature of the axial anomaly. The solutions are generally self-consistent for this version of axion electrodynamics.

In the second version, rather than starting from an effective action, the axion equations are instead obtained from a semi-classical kinetic theory as in Ref.~\cite{son2013chiral}. In contrast to the first version, this yields a nonlinear chiral mangnetic term. We found that the resulting equations generally do not admit self-consistent, physical solutions. In the case of a semi-infinite slab, no solutions exist aside from a few special cases, while for an infinite cylindrical wire, solutions exist but exhibit unphysical field divergences. These findings suggest that this version of axion electrodynamics, which has been considered in several recent experimental works, may not be self-consistent. 

The third version of axion electrydnamics we considered involves dynamical axions. That is, the axions are described by independent fields rather than by fixed background fields as in the previous two versions. We found that self-consistent solutions can be obtained only in cases where the chiral magnetic term is exactly zero, as otherwise the solution violates energy conservation. It is possible that this issue could be lifted in the case of time-dependent applied fields.

Going forward, more work needs to be done, both theoretically and experimentally, to better understand the nature of the axial anomaly in WSMs and the impact it has on the electromagnetic response of these materials. 

\section*{Acknowledgments}
E.B. acknowledges support from the National Science Foundation, grant no. DMR-1847078. The work of D.M. is supported in part by the Department
of Energy (under DOE grant number DE-SC0020262) and
the Julian Schwinger Foundation. D.M. is also grateful to Perimeter
Institute for hospitality and support. J. J. H.
acknowledges support by the U.S. Department of Energy, Office of Basic
Energy Sciences, Division of Materials Sciences and Engineering (under
Award No. DE-FG02-08ER46532).

\appendix

\section{Conductivity}\label{app:cond}
From the calculation by Ref.~\cite{throckmorton2015many}, we write down the conductivity
\begin{align}
\sigma(\omega)=\frac{1}{i\omega+\frac{1}{\tau}}\cross\frac{v_F^2e^2g}{3\pi^2(\hbar v_F)^3}\int_0^\infty {d\varepsilon \varepsilon^2(-\frac{\partial f^0(\varepsilon,T)}{\partial\varepsilon})}.
\end{align}
Here $e$, $v_F$, $g$ and $\tau$ are the electron charge, Fermi velocity, light-matter coupling and scattering time, respectively. $f^0(\varepsilon,T)$ is the Fermi-Dirac distribution. The integral above leads to a constant $\Gamma(T)$ decided by the temperature. Considering the limit $\omega\rightarrow 0$ and $T\rightarrow 0$, denoting $\Gamma_0=\Gamma(0)$ and $\sigma_0=\sigma(0)$, we have 
\begin{align}
\sigma_0=\frac{v_F^2e^2g\tau\Gamma_0}{3\pi^2(\hbar v_F)^3}.
\end{align}
Now we calculate this $\Gamma_0$; since
\begin{align}
f^0(\varepsilon)=\frac{1}{e^{\frac{\varepsilon-\mu}{k_B T}}+1},
\end{align}
we integrate by parts
\begin{align}
&\lim_{T\rightarrow 0}\int_0^\infty {d\varepsilon \varepsilon^2(-\frac{\partial f^0(\varepsilon,T)}{\partial\varepsilon})}\nonumber\\
=&-\lim_{T\rightarrow 0}\varepsilon^2f^0(\varepsilon,T)|_0^\infty+\lim_{T\rightarrow 0}\int_0^\infty{2\varepsilon f^0(\varepsilon,T)d\varepsilon}\nonumber\\
=&2\lim_{T\rightarrow 0}\int_0^\infty{\frac{\varepsilon d\varepsilon}{e^{\frac{\varepsilon-\mu}{k_B T}}+1}}=\mu^2=\varepsilon_F^2=\Gamma_0.
\end{align}
Therefore, we obtain
\begin{align}
\sigma_0=\frac{e^2g\tau\varepsilon_F^2}{3\pi^2\hbar^3v_F}=\frac{e^2g\tau k_F^2v_F^2}{3\pi^2\hbar^3v_F}=\frac{e^2g\tau k_F^2v_F}{3\pi^2\hbar^3}.
\end{align}
Meanwhile the zero-temperature carrier density is \cite{throckmorton2015many}
\begin{align}
n=\frac{gk_F^3}{6\pi^2}.
\end{align}
Therefore the conductivity and the carrier density have the relation $\sigma_0\propto n^{\frac{2}{3}}$.

\section{Non-dynamical axions and linear chiral magnetic term}\label{app:NDTD}
Here, we show in detail how we obtain the solutions described in Sec.~\ref{sec:ND axion: TD}. Starting with Eqs.~\eqref{eq: derivation E1} -~\eqref{eq: theta form}, we consider both vanishing current and non-zero current cases. Based on Ohm's law, we assume
\begin{align}
\vec{j}=\sigma_0\vec{E},
\end{align}
where the conductivity $\sigma_0$ is calculated in App.~\ref{app:cond}. In WSMs, $n$ can be very low. When this happens, the Ohmic conductance can be ignored, and we can set $\vec{j}=0$ in the axion equations. In other cases when $n$ is large enough, the conductivity cannot be ignored, and thus $\vec{j}\ne0$.

\subsection{Zero current case: $\vec{j}=0$}\label{app:NDTD1}
Now we set $\rho=0$, $\vec{j}=0$ and $\vec{E}(\vec{r},t)=e^{i\omega t}\vec{E}(\vec{r})$, $\vec{B}(\vec{r},t)=e^{i\omega t}\vec{B}(\vec{r})$. Since the EM fields are necessarily real, considering the time derivative relations, if we focus on the real part of $e^{i\omega t}$ in $\vec{E}(\vec{r},t)$, we should take the real part of $\vec{E}(\vec{r})$ as well, and correspondingly we should take the imaginary part of $e^{i\omega t}$ and $\vec{B}(\vec{r})$ in $\vec{B}(\vec{r},t)$. Thus, we have 
\begin{align}
\vec{\nabla}\cdot\vec{E}&=-\mu_0c\kappa\Delta\vec{p}\cdot\vec{B}\label{eq:TDS_E1}\\
\vec{\nabla}\cross\vec{E}&=-i\omega\vec{B} \label{eq:TDS_E2}\\
\vec{\nabla}\cdot\vec{B}&=0\label{eq:TDS_B1}\\
\vec{\nabla}\cross\vec{B}&=i\frac{\omega}{c^2}\vec{E}+\frac{\mu_0\kappa}{c}(-\Delta\varepsilon\vec{B}+\Delta\vec{p}\cross\vec{E}).\label{eq:TDS_B2}
\end{align}

We consider a semi-infinite slab with WSM filling $z\geq0$. According to the symmetry of this setup, one should expect the fields to only depend on $z$. First let us consider the fields outside the WSM. In this region ($z<0$), we have
\begin{align}
\vec{\nabla}\cdot\vec{E}&=0\\
\vec{\nabla}\cross\vec{E}&=-i\omega\vec{B}\\
\vec{\nabla}\cdot\vec{B}&=0\\
\vec{\nabla}\cross\vec{B}&=i\frac{\omega}{c^2}\vec{E}
\end{align}
Thus, we have
\begin{align}
\partial_z E_z&=0\\
\partial_z E_x\hat{y}-\partial_z E_y\hat{x}&=-i\omega\vec{B}\\
\partial_z B_z&=0\\
\partial_z B_x\hat{y}-\partial_z B_y\hat{x}&=\frac{1}{c^2}i\omega\vec{E}.
\end{align}
The solution is
\begin{align}
E_x&=E_{x,1}e^{i\frac{\omega}{c}z}+E_{x,2}e^{-i\frac{\omega}{c}z}\\
E_y&=E_{y,1}e^{i\frac{\omega}{c}z}+E_{y,2}e^{-i\frac{\omega}{c}z}\\
B_x&=\frac{1}{c}E_{y,1}e^{i\frac{\omega}{c}z}-\frac{1}{c}E_{y,2}e^{-i\frac{\omega}{c}z}\\
B_y&=-\frac{1}{c}E_{x,1}e^{i\frac{\omega}{c}z}+\frac{1}{c}E_{x,2}e^{-i\frac{\omega}{c}z}
\end{align}
For the explicit example discussed in Sec.~\ref{sec:ND axion: TD}, we choose $E_z=0=B_z$, $E_{x,1}=E_{x,2}=\frac{1}{2}E_{x,0}$, and $E_{y,1}=E_{y,2}=\frac{1}{2}E_{y,0}$, and so we have
\begin{align}
E_x&=E_{x,0}\cos{\frac{\omega}{c}z}\label{eq:TDS_vac_E1}\\
E_y&=E_{y,0}\cos{\frac{\omega}{c}z}\label{eq:TDS_vac_E2}\\
B_x&=\frac{i}{c}E_{y,0}\sin{\frac{\omega}{c}z}\label{eq:TDS_vac_B1}\\
B_y&=-\frac{i}{c}E_{x,0}\sin{\frac{\omega}{c}z}\label{eq:TDS_vac_B2}.
\end{align}

To obtain the boundary conditions at the surface of the WSM ($z=0$), one can integrate over an infinitely small volume or area that overlaps the boundary. This leads to the requirement that the fields be continuous at the boundary, as we now show. Eq.~\eqref{eq:TDS_E1} gives the integral
\begin{align}
&\lim_{V\rightarrow 0}\int{\vec{\nabla}\cdot\vec{E}dV}=\lim_{V\rightarrow 0}\oint{\vec{E}\cdot d\vec{S}}\nonumber\\
=&-\lim_{V\rightarrow 0}\kappa c^2\int{\Delta\vec{p}\cdot\vec{B}dV}=0\\
&E_z^{in}|_{z=0}=E_z^{out}|_{z=0}.
\end{align}
By doing the loop line integral, Eq.~\eqref{eq:TDS_E2} gives
\begin{align}
&\lim_{S\rightarrow 0}\int{(\vec{\nabla}\cross\vec{E})\cdot d\vec{S}}=\lim_{S\rightarrow 0}\oint{\vec{E}}\cdot d\vec{l}\nonumber\\
=&-i\omega\lim_{S\rightarrow 0}\int{\vec{B}\cdot d\vec{S}}=0\\
&E_x^{in}=E_x^{out}\\
&E_y^{in}=E_y^{out}.
\end{align}
Again, by doing the volume and loop line integrals of Eq.~\eqref{eq:TDS_B1} and Eq.~\eqref{eq:TDS_B2} respectively, we obtain
\begin{align}
B_z^{in}=&B_z^{out}\\
B_x^{in}=&B_x^{out}\\
B_y^{in}=&B_y^{out}.
\end{align}
Therefore, for the particular example of Eqs.~\eqref{eq:TDS_vac_E1}-\eqref{eq:TDS_vac_B2}, at the boundary $z\rightarrow 0$ one has
\begin{align}
E_x(0)&=E_{x,0}=E_x^{out}\\
E_y(0)&=E_{y,0}=E_y^{out}\\
B_x(0)&=0\\
B_y(0)&=0\\
E_z(0)&=0\\
B_z(0)&=0.
\end{align}

Inside the WSM, according to Eqs.~\eqref{eq:TDS_E1} -~\eqref{eq:TDS_B2}, we have
\begin{align}
\partial_zE_z=&-\mu_0c\kappa(\Delta p_xB_x+\Delta p_yB_y+\Delta p_zB_z)\label{eq:TDS_E3}\\
\partial_zE_x\hat{y}-\partial_zE_y\hat{x}=&-i\omega\vec{B}\label{eq:TDS_E4}\\
\partial_zB_z=&0\label{eq:TDS_B3}\\
\partial_zB_x\hat{y}-\partial_zB_y\hat{x}=&\frac{1}{c^2}i\omega\vec{E}-\frac{\mu_0\kappa}{c}\Delta\varepsilon\vec{B}+\frac{\mu_0\kappa}{c}[(\Delta p_yE_z\nonumber\\
&-\Delta p_zE_y)\hat{x}+(\Delta p_zE_x-\Delta p_xE_z)\hat{y}\nonumber\\
&+(\Delta p_xE_y-\Delta p_yE_x)\hat{z}].\label{eq:TDS_B4}
\end{align}
Eq.~\eqref{eq:TDS_E4} gives
\begin{align}
B_z&=0\\
B_y&=\frac{i}{\omega}\partial_z E_x\\
B_x&=-\frac{i}{\omega}\partial_z E_y,
\end{align}
which also satisfies Eq.~\eqref{eq:TDS_B3}. Now we take a look at the $z$ component of Eq.~\eqref{eq:TDS_B4}:
\begin{align}
E_z=\mu_0\kappa c\frac{i}{\omega}(\Delta p_xE_y-\Delta p_yE_x).\label{eq:TDS_B4_3}
\end{align}
Taking the derivative with respect to $z$ on both sides, we obtain
\begin{align}
\partial_zE_z=\mu_0\kappa c\frac{i}{\omega}(\Delta p_x\partial_zE_y-\Delta p_y\partial_zE_x).
\end{align}
Replacing the electric field derivatives by  $B$ field components, we have
\begin{align}
\partial_zE_z=-\mu_0\kappa c(\Delta p_xB_x+\Delta p_yB_y),
\end{align}
which is exactly Eq.~\eqref{eq:TDS_E3}. This is consistent with the interpretation of Eq.~\eqref{eq:TDS_E3} as a boundary condition in time, as discussed in Ref.~\cite{Griffiths:1492149,jackson_classical_1999}. Plugging Eq.~\eqref{eq:TDS_B4_3} into Eq.~\eqref{eq:TDS_B4} allows us to reduce the number of variables down to only $E_x$ and $E_y$. Thus we have
\begin{align}
\partial_z^2E_y+\frac{\omega^2}{c^2}E_y-\frac{\mu_0\kappa}{c}\Delta\varepsilon\partial_zE_x\nonumber\\
-\mu_0^2\kappa^2\Delta p_x(\Delta p_xE_y-\Delta p_yE_x)-i\omega\frac{\mu_0\kappa}{c}\Delta p_zE_x&=0 \label{eq:reduced1}\\
\partial_z^2E_x+\frac{\omega^2}{c^2}E_x+\frac{\mu_0\kappa}{c}\Delta\varepsilon\partial_zE_y\nonumber\\
+\mu_0^2\kappa^2\Delta p_y(\Delta p_xE_y-\Delta p_yE_x)+i\omega\frac{\mu_0\kappa}{c}\Delta p_zE_y&=0. \label{eq:reduced2}
\end{align}
Now we can use the operator method to solve these two equations. First we replace the derivatives with a parameter: $\partial_z=d$, which converts the differential equations into algebraic equations that can then be cast into a vanishing determinant condition:
\begin{widetext}
\begin{equation}
      \begin{cases}
      (d^2+\frac{\omega^2}{c^2}-\mu_0^2\kappa^2\Delta p_x^2)E_y-(\Delta\varepsilon\frac{\mu_0\kappa}{c} d-\mu_0^2\kappa^2\Delta p_x\Delta p_y+i\omega\frac{\mu_0\kappa}{c}\Delta p_z)E_x=0\\
      (\Delta\varepsilon\frac{\mu_0\kappa}{c} d+\mu_0^2\kappa^2\Delta p_x\Delta p_y+i\omega\frac{\mu_0\kappa}{c}\Delta p_z)E_y+(d^2+\frac{\omega^2}{c^2}-\mu_0\kappa^2\Delta p_y^2)E_x=0
      \end{cases}\label{eq:slab final2}
\end{equation}
\begin{equation}
\left|
\begin{array}{cccc}
d^2+\frac{\omega^2}{c^2}-\mu_0^2\kappa^2\Delta p_x^2&    -(\Delta\varepsilon\frac{\mu_0\kappa}{c} d-\mu_0^2\kappa^2\Delta p_x\Delta p_y+i\omega\frac{\mu_0\kappa}{c}\Delta p_z)\\
\Delta\varepsilon\frac{\mu_0\kappa}{c} d+\mu_0^2\kappa^2\Delta p_x\Delta p_y+i\omega\frac{\mu_0\kappa}{c}\Delta p_z&    d^2+\frac{\omega^2}{c^2}-\mu_0\kappa^2\Delta p_y^2
\end{array}
\right|=0,
\end{equation}
\end{widetext}
which gives
\begin{align}
&d^4+(2\frac{\omega^2}{c^2}-\mu_0^2\kappa^2(\Delta p_x^2+\Delta p_y^2)+\Delta\varepsilon^2\frac{\mu_0^2\kappa^2}{c^2})d^2\nonumber\\
&+2i\omega\Delta\varepsilon\frac{\mu_0^2\kappa^2}{c^2}\Delta p_zd+\frac{\omega^4}{c^4}\nonumber\\
&-\frac{\mu_0^2\kappa^2\omega^2}{c^2}(\Delta p_x^2+\Delta p_y^2+\Delta p_z^2)=0
\end{align}
The four roots $d_1,d_2,d_3,d_4$ of this characteristic equation are generically all different. The solution for the transverse electric field components can then be expressed in terms of these roots:
\begin{equation}
     E_y=\sum_{i=1}^4a_ie^{d_iz},\qquad
     E_x=\sum_{i=1}^4b_ie^{d_iz}.
\label{eq:generalsolution}
\end{equation}
One can put these expressions back into Eq.~\eqref{eq:slab final2} to reduce the 8 unknown coefficients $a_i$ and $b_i$ to 4, the rest of which are determined by boundary conditions.

Now if we assume the momentum separation is only along the $z$ direction, i.e. $\Delta p_x=\Delta p_y=0$, we have $E_z=0$ and the two equations become
\begin{align}
\partial_z^2E_y+\frac{\omega^2}{c^2}E_y-\frac{\mu_0\kappa}{c}\Delta\varepsilon\partial_zE_x-i\omega\frac{\mu_0\kappa}{c}\Delta p_zE_x&=0\\
\partial_z^2E_x+\frac{\omega^2}{c^2}E_x+\frac{\mu_0\kappa}{c}\Delta\varepsilon\partial_zE_y+i\omega\frac{\mu_0\kappa}{c}\Delta p_zE_y&=0.
\end{align}
Thus the operator equation becomes
\begin{align}
&d^4+(2\frac{\omega^2}{c^2}+\Delta\varepsilon^2\frac{\mu_0^2\kappa^2}{c^2})d^2+2i\omega\Delta\varepsilon
\frac{\mu_0^2\kappa^2}{c^2}\Delta p_zd\nonumber\\ &+\frac{\omega^4}{c^4}-\frac{\mu_0^2\kappa^2\omega^2}{c^2}\Delta p_z^2=0.
\end{align}

\subsection{Non-zero case: $\vec{j}\ne 0$}\label{app:NDTD2}
Here we set $\vec{j}=\sigma_0\vec{E}$ and $\vec{E}(\vec{r},t)=e^{i\omega t}\vec{E}(\vec{r})$, $\vec{B}(\vec{r},t)=e^{i\omega t}\vec{B}(\vec{r})$. Therefore the current can also be separated into temporal and spatial parts: $\vec{j}(\vec{r},t)=e^{i\omega t}\vec{j}(\vec{r})$. As for $\rho$, one will find this cannot be zero, as we will see later. Again, since the EM fields are necessarily real, considering the time derivative relations, if we focus on the real part of $e^{i\omega t}$ in $\vec{E}(\vec{r},t)$, we should take the real part of $\vec{E}(\vec{r})$ as well, and correspondingly we should take the imaginary part of $e^{i\omega t}$ and $\vec{B}(\vec{r})$ in $\vec{B}(\vec{r},t)$. Thus, we have 
\begin{align}
\vec{\nabla}\cdot\vec{E}&=\frac{\rho}{\varepsilon_0}-\mu_0c\kappa\Delta\vec{p}\cdot\vec{B}\label{eq:omTDS_E1}\\
\vec{\nabla}\cross\vec{E}&=-i\omega\vec{B} \label{eq:omTDS_E2}\\
\vec{\nabla}\cdot\vec{B}&=0\label{eq:omTDS_B1}\\
\vec{\nabla}\cross\vec{B}&=i\frac{\omega}{c^2}\vec{E}+\mu_0\sigma_0\vec{E}+\frac{\mu_0\kappa}{c}(-\Delta\varepsilon\vec{B}+\Delta\vec{p}\cross\vec{E}).\label{eq:omTDS_B2}
\end{align}
The boundary conditions and outside fields are the same as when $\vec{j}=0$ since charges and currents are in the bulk instead of on the surface. Therefore, at the boundary $z\rightarrow 0$ one has
\begin{align}
E_x(0)&=E_{x,0}=E_x^{out}\\
E_y(0)&=E_{y,0}=E_y^{out}\\
B_x(0)&=0\\
B_y(0)&=0\\
E_z(0)&=0\\
B_z(0)&=0.
\end{align}
Now we have
\begin{align}
\partial_zE_z=&\frac{\rho}{\varepsilon_0}-\mu_0c\kappa(\Delta p_xB_x+\Delta p_yB_y+\Delta p_zB_z)\label{eq:omTDS_E3}\\
\partial_zE_x\hat{y}-\partial_zE_y\hat{x}=&-i\omega\vec{B}\label{eq:omTDS_E4}\\
\partial_zB_z=&0\label{eq:omTDS_B3}\\
\partial_zB_x\hat{y}-\partial_zB_y\hat{x}=&\frac{1}{c^2}i\omega\vec{E}+\mu_0\sigma_0\vec{E}-\frac{\mu_0\kappa}{c}\Delta\varepsilon\vec{B}\nonumber\\&+\frac{\mu_0\kappa}{c}[(\Delta p_yE_z-\Delta p_zE_y)\hat{x}\nonumber\\
&+(\Delta p_zE_x-\Delta p_xE_z)\hat{y}\nonumber\\
&+(\Delta p_xE_y-\Delta p_yE_x)\hat{z}].\label{eq:omTDS_B4}
\end{align}
Eq.~\eqref{eq:omTDS_E4} leads
\begin{align}
B_z&=0\\
B_y&=\frac{i}{\omega}\partial_z E_x\label{eq:omTDS_BDE_1}\\
B_x&=-\frac{i}{\omega}\partial_z E_y\label{eq:omTDS_BDE_2},
\end{align}
Now let us look at the $z$ component of Eq.~\eqref{eq:omTDS_B4}:
\begin{align}
E_z=i\frac{\mu_0\sigma_0 c^2}{\omega}E_z+\mu_0\kappa c\frac{i}{\omega}(\Delta p_xE_y-\Delta p_yE_x).\label{eq:omTDS_B4_3}
\end{align}
Taking the derivative with respect to $z$ on both sides, we obtain
\begin{align}
\partial_zE_z=i\frac{\mu_0\sigma_0 c^2}{\omega}\partial_zE_z+\mu_0\kappa c\frac{i}{\omega}(\Delta p_x\partial_zE_y-\Delta p_y\partial_zE_x).
\end{align}
Replacing the electric field derivatives by  $B$ field components, we have
\begin{align}
\partial_zE_z=i\frac{\mu_0\sigma_0 c^2}{\omega}\partial_zE_z-\mu_0\kappa c(\Delta p_xB_x+\Delta p_yB_y).
\end{align}
Comparing this equation to Eq.~\eqref{eq:omTDS_E3}, we see that self-consistency requires
\begin{align}
\frac{\rho}{\varepsilon_0}=i\frac{\mu_0\sigma_0 c^2}{\omega}\partial_zE_z.
\end{align}
If the charge density were zero here, we would have
\begin{align}
E_z&=E_{z_0}=\Delta p_xE_y-\Delta p_yE_x\\
\partial_zE_z&=0=\Delta p_x\partial_zE_y-\Delta p_y\partial_zE_x,
\end{align}
which means $E_x$ and $E_y$ should be linearly dependent on each other. The same is true for the magnetic fields due to Eq.~\eqref{eq:omTDS_BDE_1} and ~\eqref{eq:omTDS_BDE_2}. However, we are free to choose the boundary conditions outside, and so it requires fine tuning to obey these conditions. Therefore, if we include the non-zero current, we must have non-zero net bulk charges, which means $\rho\ne 0$ as well. Another possibility is that $\Delta p_x=\Delta p_y=0$. In this special case we can have zero net bulk charge. This is also the simple case we will consider next.

In this case, we can still use our previous approach to solve these new equations. For simplicity, we still consider the momentum separation to be only along the $z$ direction, i.e. $\Delta p_x=\Delta p_y=0$, we have $E_z=0$ and the two equations become
\begin{align}
\partial_z^2E_y+\frac{\omega^2}{c^2}E_y-i\mu_0\sigma_0\omega E_y\nonumber\\
-\frac{\mu_0\kappa}{c}\Delta\varepsilon\partial_zE_x-i\omega\frac{\mu_0\kappa}{c}\Delta p_zE_x&=0\\
\partial_z^2E_x+\frac{\omega^2}{c^2}E_x-i\mu_0\sigma_0\omega E_x\nonumber\\
+\frac{\mu_0\kappa}{c}\Delta\varepsilon\partial_zE_y+i\omega\frac{\mu_0\kappa}{c}\Delta p_zE_y&=0.
\end{align}
The operator equation becomes
\begin{align}
&d^4+[2(\frac{\omega^2}{c^2}-i\mu_0\sigma_0\omega)+\Delta\varepsilon^2\frac{\mu_0^2\kappa^2}{c^2}]d^2+2i\omega\Delta\varepsilon
\frac{\mu_0^2\kappa^2}{c^2}\Delta p_zd\nonumber\\ &+(\frac{\omega^2}{c^2}-i\mu_0\sigma_0\omega)^2-\frac{\mu_0^2\kappa^2\omega^2}{c^2}\Delta p_z^2=0.
\end{align}
The remaining steps are described in the main text.

\section{Non-dynamical axions with nonlinear chiral magnetic term}\label{app:NDTI}
Here, we provide details about the solutions described in Sec.~\ref{sec:ND axion: TID}. We consider three geometries for the WSM: a semi-infinite slab, whole space, and an infinite cylindrical wire. In all cases, the starting point is a version of axion electrodynamics in which the chiral magnetic term is nonlinear:

\begin{align}
\vec{\nabla}\cdot\vec{E}&=\frac{\rho}{\varepsilon_0}-\mu_0c\kappa\Delta\vec{p}\cdot\vec{B}\label{omeq:TIND_E1}\\
\vec{\nabla}\cross\vec{E}&=0\label{omeq:TIND_E2}\\
\vec{\nabla}\cdot\vec{B}&=0\label{omeq:TIND_B1}\\
\vec{\nabla}\cross\vec{B}&=\mu_0\sigma_0\vec{E}+\mu_0\sigma_a(\vec{E}\cdot\vec{B})\vec{B}+\frac{\mu_0\kappa}{c}\Delta\vec{p}\cross\vec{E}.\label{omeq:TIND_B2}
\end{align}
First let us consider a semi-infinite slab case when $\vec{j}=\sigma_0\vec{E}\ne 0$. The WSM occupies $z\geq0$, while $z<0$ is vacuum. We assume the outside fields are in the $xy$ plane: $\vec{E}=E_{x,0}\hat{x}+E_{y,0}\hat{y}$ and $\vec{B}=B_{x,0}\hat{x}+B_{y,0}\hat{y}$. From Eq.~\eqref{omeq:TIND_E1}, we have
\begin{align}
\partial_z E_z=\frac{\rho}{\varepsilon_0}-\mu_0c\kappa(\Delta p_xB_x+\Delta p_yB_y+\Delta p_zB_z)
\end{align}
From Eq.~\eqref{omeq:TIND_E2}, we have
\begin{align}
\partial_zE_x\hat{y}-\partial_zE_y\hat{x}&=0\\
E_x=const&=E_{x,0}\\
E_y=const&=E_{y,0}
\end{align}
From Eq.~\eqref{omeq:TIND_B1}, we have
\begin{align}
B_z=const.
\end{align}
From Eq.~\eqref{omeq:TIND_B2}, we have
\begin{align}
&\partial_zB_x\hat{y}-\partial_zB_y\hat{x}\nonumber\\
=&(\mu_0\sigma_0E_x+\mu_0\sigma_aE_iB_iB_x+\frac{\mu_0\kappa}{c}\Delta p_yE_z-\frac{\mu_0\kappa}{c}\Delta p_z E_y)\hat{x}\nonumber\\
+&(\mu_0\sigma_0E_y+\mu_0\sigma_aE_iB_iB_y+\frac{\mu_0\kappa}{c}\Delta p_zE_x-\frac{\mu_0\kappa}{c}\Delta p_xE_z)\hat{y}\nonumber\\
+&(\mu_0\sigma_0E_z+\mu_0\sigma_aE_iB_iB_z+\frac{\mu_0\kappa}{c}\Delta p_xE_y-\frac{\mu_0\kappa}{c}\Delta p_yE_x)\hat{z}
\end{align}
Now we have three variables $E_z, B_x$ and $B_y$ and four equations:
\begin{align}
\partial_zE_z=&\frac{\rho}{\varepsilon_0}-\mu_0c\kappa(\Delta p_xB_x+\Delta p_yB_y)\\
\partial_zB_y=&\mu_0\sigma_0E_{x,0}-[\mu_0\sigma_a(E_{x,0}B_x+E_{y,0}B_y)B_x\nonumber\\
&+\frac{\mu_0\kappa}{c}\Delta p_yE_z-\frac{\mu_0\kappa}{c}\Delta p_z E_{y,0}]\\
\partial_zB_x=&\mu_0\sigma_0E_{y,0}+\mu_0\sigma_a(E_{x,0}B_x+E_{y,0}B_y)B_y\nonumber\\
&+\frac{\mu_0\kappa}{c}\Delta p_zE_{x,0}-\frac{\mu_0\kappa}{c}\Delta p_xE_z\\
\mu_0\sigma_0E_z=&\Delta p_yE_{x,0}-\Delta p_xE_{y,0}\label{omeq:TIND_slab_inconsis}.
\end{align}
The last equation tells us $E_z$ is a constant determined by the Weyl separations and the boundary conditions of $E_x$ and $E_y$, which implies
\begin{align}
E_z=E_{z,0}=\frac{\Delta p_yE_{x,0}-\Delta p_xE_{y,0}}{\mu_0\sigma_0}\label{omeq:TIND_slab_inconsis}.
\end{align}
However, as we know $E_{z,0}$ should be chosen freely. Thus, this fine-tuning problem leads a generic inconsistency. For simplicity, from now on, we assume that the current and charge are both zero, i.e., $\rho=0$ and $\vec{j}=0$:
\begin{align}
\vec{\nabla}\cdot\vec{E}&=-\mu_0c\kappa\Delta\vec{p}\cdot\vec{B}\label{eq:TIND_E1}\\
\vec{\nabla}\cross\vec{E}&=0\label{eq:TIND_E2}\\
\vec{\nabla}\cdot\vec{B}&=0\label{eq:TIND_B1}\\
\vec{\nabla}\cross\vec{B}&=\mu_0\sigma_a(\vec{E}\cdot\vec{B})\vec{B}+\frac{\mu_0\kappa}{c}\Delta\vec{p}\cross\vec{E}.\label{eq:TIND_B2}
\end{align}

\subsection{Semi-infinite slab case}\label{app:NDTI_slab}
We first consider a semi-infinite slab of WSM occupying $z\geq0$, while $z<0$ is vacuum. First, we assume the outside fields are in the $xy$ plane: $\vec{E}=E_{x,0}\hat{x}+E_{y,0}\hat{y}$ and $\vec{B}=B_{x,0}\hat{x}+B_{y,0}\hat{y}$. From Eq.~\eqref{eq:TIND_E1}, we have
\begin{align}
\partial_z E_z=-\mu_0c\kappa(\Delta p_xB_x+\Delta p_yB_y+\Delta p_zB_z)
\end{align}
From Eq.~\eqref{eq:TIND_E2}, we have
\begin{align}
\partial_zE_x\hat{y}-\partial_zE_y\hat{x}&=0\\
E_x=const&=E_{x,0}\\
E_y=const&=E_{y,0}
\end{align}
From Eq.~\eqref{eq:TIND_B1}, we have
\begin{align}
B_z=const.
\end{align}
From Eq.~\eqref{eq:TIND_B2}, we have
\begin{align}
&\partial_zB_x\hat{y}-\partial_zB_y\hat{x}\nonumber\\
=&(\mu_0\sigma_aE_iB_iB_x+\frac{\mu_0\kappa}{c}\Delta p_yE_z-\frac{\mu_0\kappa}{c}\Delta p_z E_y)\hat{x}\nonumber\\
&+(\mu_0\sigma_aE_iB_iB_y+\frac{\mu_0\kappa}{c}\Delta p_zE_x-\frac{\mu_0\kappa}{c}\Delta p_xE_z)\hat{y}\nonumber\\
&+(\mu_0\sigma_aE_iB_iB_z+\frac{\mu_0\kappa}{c}\Delta p_xE_y-\frac{\mu_0\kappa}{c}\Delta p_yE_x)\hat{z}
\end{align}
Now we have three variables $E_z, B_x$ and $B_y$ and four equations:
\begin{align}
\partial_zE_z=&-\mu_0c\kappa(\Delta p_xB_x+\Delta p_yB_y)\\
\partial_zB_y=&-[\mu_0\sigma_a(E_{x,0}B_x+E_{y,0}B_y)B_x\nonumber\\
&+\frac{\mu_0\kappa}{c}\Delta p_yE_z-\frac{\mu_0\kappa}{c}\Delta p_z E_{y,0}]\\
\partial_zB_x=&\mu_0\sigma_a(E_{x,0}B_x+E_{y,0}B_y)B_y\nonumber\\
&+\frac{\mu_0\kappa}{c}\Delta p_zE_{x,0}-\frac{\mu_0\kappa}{c}\Delta p_xE_z\\
\Delta p_xE_{y,0}=&\Delta p_yE_{x,0}\label{eq:TIND_slab_inconsis}.
\end{align}

Following the same logic as in Appendix~\ref{app:NDTD}, it again follows that all the fields are continuous at the boundary. Thus, Eq.~\eqref{eq:TIND_slab_inconsis} gives a strong constraint on the fields outside of the Weyl semimetal. Since the separation of two Weyl nodes is given and fixed, this is inconsistent with the free choice of the fields outside the sample.

Let us ignore this inconsistency for the time being and consider the case where the outside fields are in the $z$ direction, i.e. $\vec{E}=E_{z,0}\hat{z}$ and $\vec{B}=B_{z,0}\hat{z}$, and further assume $\Delta p_x=\Delta p_y=0$. Directly from Eq.~\eqref{eq:TIND_E1}, we have
\begin{align}
\partial_zE_z=-\mu_0c\kappa\Delta p_zB_z
\end{align}
From Eq.~\eqref{eq:TIND_E2}, we have
\begin{align}
\partial_zE_x\hat{y}-\partial_zE_y\hat{x}&=0\\
E_x=const&=E_{x,0}^{out}=0\\
E_y=const&=E_{y,0}^{out}=0
\end{align}
From Eq.~\eqref{eq:TIND_B1}, we have
\begin{align}
B_z=const=B_{z,0}
\end{align}
From Eq.~\eqref{eq:TIND_B2}, we have
\begin{align}
&\partial_zB_x\hat{y}-\partial_zB_y\hat{x}\nonumber\\
=&\mu_0\sigma_aE_zB_{z,0}B_x\hat{x}+\mu_0\sigma_aE_zB_{z,0}B_y\hat{y}+\mu_0\sigma_aE_zB_{z,0}^2\hat{z}.
\end{align}
Since we are free to choose $B_{z,0}$ (which is a component of the applied magnetic field), the $z$ component of this last equation gives
\begin{align}
E_z=0,
\end{align}
while the other two components yield
\begin{align}
B_x=&B_{x,0}=0\\
B_y=&B_{y,0}=0\\
\partial_zE_z=&0=-\mu_0c\kappa\Delta p_zB_z\\
B_z=&B_{z,0}=0.
\end{align}
This contradicts the assumption that the applied magnetic field is nonzero, $B_{z0} \neq 0$. Thus, we again arrive at an inconsistent solution.

\subsection{Whole space case}\label{app:NDTI_whole}
Here, we consider the case where the whole space is a WSM. Thus, the fields must be constant due to symmetry. We have
\begin{align}
\vec{\nabla}\cdot\vec{E}&=-\mu_0c\kappa\Delta\vec{p}\cdot\vec{B}=0\\
\vec{\nabla}\cross\vec{B}&=\mu_0\sigma_a(\vec{E}\cdot\vec{B})\vec{B}+\frac{\mu_0\kappa}{c}\Delta\vec{p}\cross\vec{E}=0.
\end{align}
The WSM has the momentum separation $\Delta\vec{p}$. We choose the direction of this vector to be $\hat{z}$, i.e. $\Delta\vec{p}=\Delta p_z \hat{z}$, and so we have $\vec{B}=(B_x,B_y,0)$ from the first equation. From the second equation, we obtain
\begin{align}
\mu_0\sigma_a(E_xB_x+E_yB_y)B_x-\frac{\mu_0\kappa}{c}\Delta p_zE_y=0\\
\mu_0\sigma_a(E_xB_x+E_yB_y)B_y+\frac{\mu_0\kappa}{c}\Delta p_zE_x=0,
\end{align}
which gives
\begin{align}
E_x&=-\frac{\mu_0\sigma_aB_xB_y-\frac{\mu_0\kappa}{c}\Delta p_z}{\mu_0\sigma_aB_x^2}E_y\\
E_x&=-\frac{\mu_0\sigma_aB_y^2}{\mu_0\sigma_aB_xB_y+\frac{\mu_0\kappa}{c}\Delta p_z}E_y.
\end{align}
So we have $E_x=E_y=0$ or
\begin{align}
\mu_0^2\sigma_a^2B_x^2B_y^2-(\frac{\mu_0\kappa}{c})^2\Delta p_z^2=\mu_0^2\sigma_a^2B_x^2B_y^2.
\end{align}
Since $\Delta p_z\neq0$, the only possibility is 
\begin{align}
E_x&=E_y=0,\\
E_z&=E_{z,0}.
\end{align}
This means the electric field can only exist along the direction of the Weyl separation, while the magnetic field must be perpendicular to this direction. Thus, the CME cannot exist in this case.

\subsection{Cylindrical wire case}\label{App_B3}
Here, we consider an infinite cylindrical wire with radius $R$ made from a WSM. The axis of the cylinder is along the $\hat{z}$ direction. Again we start with Eqs.~\eqref{eq:TIND_E1}-\eqref{eq:TIND_B2}. To maintain cylindrical symmetry, we focus on the case $\Delta\vec{p}=\Delta p_z\hat z$. Because of this symmetry, all fields should depend on $r$ only. We obtain the equations
\begin{align}
\frac{1}{r}\frac{\partial}{\partial r}(rE_r)&=-\mu_0c\kappa\Delta p_zB_z\\
\frac{\partial E_z}{\partial r}&=0\\
\frac{1}{r}\frac{\partial}{\partial r}(rE_\phi)&=0\\
\frac{1}{r}\frac{\partial}{\partial r}(rB_r)&=0
\end{align}
\begin{align}
&-\frac{\partial B_z}{\partial r}\hat{\phi}+\frac{1}{r}\frac{\partial}{\partial r}(rB_\phi)\hat{z}\nonumber\\
=&\mu_0\sigma_a(E_rB_r+E_\phi B_\phi+E_zB_z)\vec{B}\nonumber\\
&+\frac{\mu_0\kappa}{c}\Delta p_zE_r\hat{\phi}-\frac{\mu_0\kappa}{c}\Delta p_z E_\phi\hat{r}.
\end{align}
Thus we have
\begin{align}
\frac{1}{r}\frac{\partial}{\partial r}(rE_r)+\mu_0c\kappa\Delta p_zB_z&=0\label{eq:TIND_cylinder_E3}\\
E_z=E_{z,0}&\\
E_\phi=\frac{C_1}{r}&=0\\
B_r=\frac{C_2}{r}&=0\\
\frac{\partial B_z}{\partial r}+\mu_0\sigma_aE_{z,0}B_zB_\phi+\frac{\mu_0\kappa}{c}\Delta p_zE_r&=0\\
-\frac{1}{r}\frac{\partial}{\partial r}(rB_\phi)+\mu_0\sigma_aE_{z,0}B_z^2&=0.
\end{align}
Choosing $C_1=C_2=0$ prevents some of the field components from becoming singular at $r=0$. The continuity of the fields across the WSM surface can again be established by performing volume or area integrations, as we showed for the semi-infinite slab geometry. In the case of the cylindrical wire, the same analysis yields the following continuity conditions:
\begin{align}
E_r^{in}|_{r=R}&=E_r^{out}|_{r=R}\\
E_z^{in}&=E_z^{out}=E_{z,0}\\
E_{\phi}^{in}&=E_{\phi}^{out}=0\\
B_r^{in}&=B_r^{out}=0\\
B_{\phi}^{in}|_{r=R}&=B_{\phi}^{out}|_{r=R}\\
B_{z}^{in}|_{r=R}&=B_{z}^{out}|_{r=R}.
\end{align}
In summary, all fields are continuous across the boundary.

\section{Dynamical axions}\label{app:DA}
Here, we show details of the solutions for dynamical axions obtained in Sec.~\ref{sec:D axion}. We consider a semi-infinite slab of WSM occupying the half-space $z\ge0$. We consider two cases: one in which the applied fields are orthogonal to the surface, and one in which the fields are parallel to the surface. The solution details for both cases are given below. In both cases, all the fields are continuous across the surface, as follows from an analysis similar to the one we performed for the other two versions of axion electrodynamics considered in this work.

\subsection{$\vec{E}, \vec{B} \parallel \hat{z}$ outside of the WSM}\label{app: case A}
From the main text Eqs.~\eqref{eq:case7_const_E_x} -~\eqref{eq:case7_fz}, we have
\begin{align}
E_{x,0}=E_{y,0}=0\\
f_{t,0}=0.
\end{align}
Replcing $\partial_z$ with operator $P$, we have four variables and four equations:
\begin{align}
f_{x,0}B_x+f_{y,0}B_y+\frac{1}{\mu_0c\kappa}PE_z+B_{z,0}f_z=&0\\
PB_x+\frac{\mu_0\kappa}{c}f_{x,0}E_z=&0\\
PB_y+\frac{\mu_0\kappa}{c}f_{y,0}E_z=&0\\
\frac{\kappa}{\kappa_0c}B_{z,0}E_z+Pf_z=&0
\end{align}
The determinant is then
\begin{align}
{\left|
\begin{array}{cccc}
f_{x,0}&    f_{y,0}&    \frac{1}{\mu_0c\kappa}P&    B_{z,0}\\
P&    0&    \frac{\mu_0\kappa}{c}f_{x,0}&    0\\
0&    P&    \frac{\mu_0\kappa}{c}f_{y,0}&    0\\
0&    0&    \frac{\kappa}{\kappa_0c}B_{z,0}&    P
\end{array}
\right|}=0.
\end{align}
The operator equation reads
\begin{align}
P^2(P^2-D^2)=0,
\end{align}
where
\begin{align}
D^2=\frac{\kappa^2\mu_0[B_{z,0}^2+(f_{x,0}^2+f_{y,0}^2)\kappa_0\mu_0]}{\kappa_0}.
\end{align}
This means we have four roots for P:
\begin{align}
P_1=P_2&=0\\
P_3&=D\\
P_4&=-D
\end{align}
The solutions of the ODEs are of the form
\begin{align}
x_i=a_ie^{Dz}+b_ie^{-Dz}+c_iz+d_i,
\end{align}
where we associate the indices to the fields as follows:
$B_x\to1$, $B_y\to2$, $E_z\to3$, $f_z\to4$. Putting the general forms of the solutions back into the 4 equations, we obtain
\begin{align}
f_{x,0}a_1+f_{y,0}a_2+\frac{1}{\mu_0c\kappa}D a_3+B_{z,0}a_4=&0\\
D a_1+\frac{\mu_0\kappa}{c}f_{x,0}a_3=&0\\
D a_2+\frac{\mu_0\kappa}{c}f_{y,0}a_3=&0\\
\frac{\kappa}{\kappa_0c}B_{z,0}a_3+D a_4=&0
\end{align}
\begin{align}
f_{x,0}b_1+f_{y,0}b_2-\frac{1}{\mu_0c\kappa}D b_3+B_{z,0}b_4=&0\\
-D b_1+\frac{\mu_0\kappa}{c}f_{x,0}b_3=&0\\
-D b_2+\frac{\mu_0\kappa}{c}f_{y,0}b_3=&0\\
\frac{\kappa}{\kappa_0c}B_{z,0}b_3-D b_4=&0
\end{align}
\begin{align}
f_{x,0}c_1+f_{y,0}c_2+B_{z,0}c_4=&0\\
\frac{\mu_0\kappa}{c}f_{x,0}c_3=&0\\
\frac{\mu_0\kappa}{c}f_{y,0}c_3=&0\\
\frac{\kappa}{\kappa_0c}B_{z,0}c_3=&0
\end{align}
\begin{align}
f_{x,0}d_1+f_{y,0}d_2+\frac{1}{\mu_0c\kappa}c_3+B_{z,0}d_4=&0\\
c_1+\frac{\mu_0\kappa}{c}f_{x,0}d_3=&0\\
c_2+\frac{\mu_0\kappa}{c}f_{y,0}d_3=&0\\
\frac{\kappa}{\kappa_0c}B_{z,0}d_3+c_4=&0
\end{align}

\begin{widetext}
Thus we can fix many coefficients based on the equations above
\begin{align}
a_2&=a_1\frac{f_{y,0}}{f_{x,0}}\\
a_3&=a_1\frac{c\sqrt{B_{z,0}^2+(f_{x,0}^2+f_{y,0}^2)\kappa_0\mu_0}}{f_{x,0}\sqrt{\kappa_0\mu_0}}\\
a_4&=a_1\frac{B_{z,0}}{f_{x,0}\kappa_0\mu_0}\\
b_2&=b_1\frac{f_{y,0}}{f_{x,0}}\\
b_3&=b_1\frac{c\sqrt{B_{z,0}^2+(f_{x,0}^2+f_{y,0}^2)\kappa_0\mu_0}}{f_{x,0}\sqrt{\kappa_0\mu_0}}\\
b_4&=b_1\frac{B_{z,0}}{f_{x,0}\kappa_0\mu_0}\\
c_1&=c_2=c_3=c_4=0\\
d_3&=0\\
d_4&=-\frac{d_1f_{x,0}+d_2f_{y,0}}{B_{z,0}},
\end{align}
where $a_1,b_1,d_1,d_2$ are determined by the boundary conditions of the four fields.

If we set $B_x(0)=B_y(0)=0$, $E_z(0)=E_{z,0}$ and $f_z(0)=0$, we obtain
\begin{align}
a_1&=-b_1=-\frac{E_{z,0}f_{x,0}}{2c\sqrt{f_{x,0}^2+f_{y,0}^2+\frac{B_{z,0}^2}{\kappa_0\mu_0}}}\\
d_1&=d_2=0.
\end{align}
And the solutions of the unknown fields are
\begin{align}
B_x&=-\frac{E_{z,0}f_{x,0}\kappa\mu_0\sinh(D z)}{c D}\\
B_y&=-\frac{E_{z,0}f_{y,0}\kappa\mu_0\sinh(D z)}{c D}\\
E_z&=E_{z,0}\cosh{D z}\\
f_z&=-\frac{B_{z,0}E_{z,0}\kappa\sinh{D z}}{c\kappa_0 D}
\end{align}

If we set $B_x(0)=B_y(0)=0$, $E_z(0)=E_{z,0}$ and $f_z(0)=f_{z,0}$, we obtain
\begin{align}
a_1&=-\frac{f_{x,0}\sqrt{\kappa_0\mu_0}[B_{z,0}^2E_{z,0}+E_{z,0}(f_{x,0}^2+f_{y,0}^2)\kappa_0\mu_0-B_{z,0}c f_{z,0}\sqrt{\kappa_0\mu_0}\sqrt{B_{z,0}^2+(f_{x,0}^2+f_{y,0}^2)\kappa_0\mu_0}]}{2c[B_{z,0}^2+(f_{x,0}^2+f_{y,0}^2)\kappa_0\mu_0]^{\frac{3}{2}}}\\
b_1&=\frac{f_{x,0}\sqrt{\kappa_0\mu_0}[B_{z,0}^2E_{z,0}+E_{z,0}(f_{x,0}^2+f_{y,0}^2)\kappa_0\mu_0+B_{z,0}c f_{z,0}\sqrt{\kappa_0\mu_0}\sqrt{B_{z,0}^2+(f_{x,0}^2+f_{y,0}^2)\kappa_0\mu_0}]}{2c[B_{z,0}^2+(f_{x,0}^2+f_{y,0}^2)\kappa_0\mu_0]^{\frac{3}{2}}}\\
d_1&=-\frac{B_{z,0}f_{x,0}f_{z,0}\kappa_0\mu_0}{B_{z,0}^2+(f_{x,0}^2+f_{y,0}^2)\kappa_0\mu_0}\\
d_2&=-\frac{B_{z,0}f_{y,0}f_{z,0}\kappa_0\mu_0}{B_{z,0}^2+(f_{x,0}^2+f_{y,0}^2)\kappa_0\mu_0}
\end{align}
And the solutions of the unknown fields are
\begin{align}
B_x&=\frac{f_{x,0}\kappa\mu_0[B_{z,0}c f_{z,0}\kappa\mu_0(-1+\cosh{D z})-D E_{z,0}\sinh{D z}]}{c D^2}\\
B_y&=\frac{f_{y,0}\kappa\mu_0[B_{z,0}c f_{z,0}\kappa\mu_0(-1+\cosh{D z})-D E_{z,0}\sinh{D z}]}{c D^2}\\
E_z&=E_{z,0}\cosh{D z}-\frac{B_{z,0}c f_{z,0}\kappa\mu_0}{D}\sinh{D z}\\
f_z&=\frac{c f_{z,0}(\kappa_0 D^2-B_{z,0}^2\kappa^2\mu_0+B_{z,0}^2\kappa^2\mu_0\cosh{D z})-B_{z,0}E_{z,0}\kappa D\sinh{D z}}{c\kappa_0 D^2}.
\end{align}
\end{widetext}

\subsection{$\vec{E}, \vec{B} \parallel \hat{x}$ outside of the WSM.}\label{app: case B}

Now we consider the case where the fields outside the slab are parallel to the surface. Setting $B_z=B_{z,0}=0$ and $E_{y,0}=0$, we immediately find that $f_{t,0}$, $f_{x,0}$ are free parameters, and
\begin{align}
f_{y,0}=&0\\
\partial_zE_z=&-\mu_0c\kappa f_{x,0}B_x\\
\partial_zB_x=&\frac{\mu_0\kappa}{c}(f_{t,0}B_y-f_{x,0}E_z+E_{x,0}f_z)\\
\partial_zB_y=&-\frac{\mu_0\kappa}{c}f_{t,0}B_x\\
\partial_zf_z=&-\frac{\kappa}{\kappa_0c}E_{x,0}B_x.
\end{align}
Replacing $\partial_z$ with operator $P$, again, we have four variables and four equations:
\begin{align}
-\frac{c}{\mu_0\kappa}PB_x+f_{t,0}B_y-f_{x,0}E_z+E_{x,0}f_z=&0\\
f_{t,0}B_x+\frac{c}{\mu_0\kappa}PB_y=&0\\
f_{x,0}B_x+\frac{1}{\mu_0c\kappa}PE_z=&0\\
E_{x,0}B_x+\frac{\kappa_0c}{\kappa}Pf_z=&0.
\end{align}
The determinant is then
\begin{align}
{\left|
\begin{array}{cccc}
-\frac{c}{\mu_0\kappa}P&    f_{t,0}&    -f_{x,0}&    E_{x,0}\\
f_{t,0}&    \frac{c}{\mu_0\kappa}P&    0&    0\\
f_{x,0}&    0&    \frac{1}{\mu_0c\kappa}P&    0\\
E_{x,0}&    0&    0&    \frac{\kappa_0c}{\kappa}P
\end{array}
\right|}=0.
\end{align}
The operator equation reads
\begin{align}
P^2(P^2-D^2)=0,
\end{align}
where
\begin{align}
D^2=\frac{-\kappa^2\mu_0[E_{x,0}^2+\kappa_0\mu_0(f_{t,0}^2-c^2f_{x,0}^2)]}{c^2\kappa_0}
\end{align}
This means we have four roots for P:
\begin{align}
P_1=P_2&=0\\
P_3&=D\\
P_4&=-D
\end{align}
The solutions of the ODEs have the general form
\begin{align}
x_i=a_ie^{Dz}+b_ie^{-Dz}+c_iz+d_i,
\end{align}
where the indices are associated with the field components according to
$B_x\to1$, $B_y\to2$, $E_z\to3$, $f_z\to4$. Plugging the general forms of the solutions into the 4 equations, we obtain
\begin{align}
-\frac{c}{\mu_0\kappa}D a_1+f_{t,0}a_2-f_{x,0}a_3+E_{x,0}a_4=&0\\
f_{t,0}a_1+\frac{c}{\mu_0\kappa}D a_2=&0\\
f_{x,0}a_1+\frac{D}{\mu_0c\kappa}a_3=&0\\
E_{x,0}a_1+\frac{\kappa_0c}{\kappa}D a_4=&0.
\end{align}
\begin{align}
\frac{c}{\mu_0\kappa}D b_1+f_{t,0}b_2-f_{x,0}b_3+E_{x,0}b_4=&0\\
f_{t,0}b_1-\frac{c}{\mu_0\kappa}D b_2=&0\\
f_{x,0}b_1-\frac{D}{\mu_0c\kappa}b_3=&0\\
E_{x,0}b_1-\frac{\kappa_0c}{\kappa}D b_4=&0.
\end{align}
\begin{align}
f_{t,0}c_2-f_{x,0}c_3+E_{x,0}c_4=&0\\
f_{t,0}c_1=&0\\
f_{x,0}c_1=&0\\
E_{x,0}c_1=&0.
\end{align}
\begin{align}
-\frac{c}{\mu_0\kappa}c_1+f_{t,0}d_2-f_{x,0}d_3+E_{x,0}d_4=&0\\
f_{t,0}d_1+\frac{c}{\mu_0\kappa}c_2=&0\\
f_{x,0}d_1+\frac{1}{\mu_0c\kappa}c_3=&0\\
E_{x,0}d_1+\frac{\kappa_0c}{\kappa}c_4=&0.
\end{align}
Solving these equations we obtain
\begin{align}
a_2=&i a_1\frac{f_{t,0}\sqrt{\kappa_0\mu_0}}{\sqrt{E_{x,0}^2+f_{t,0}^2\kappa_0\mu_0-c^2f_{x,0}^2\kappa_0\mu_0}}\\
a_3=&i a_1\frac{c^2f_{x,0}\sqrt{\kappa_0\mu_0}}{\sqrt{E_{x,0}^2+f_{t,0}^2\kappa_0\mu_0-c^2f_{x,0}^2\kappa_0\mu_0}}\\
a_4=&i a_1\frac{E_{x,0}}{\sqrt{\kappa_0\mu_0}\sqrt{E_{x,0}^2+f_{t,0}^2\kappa_0\mu_0-c^2f_{x,0}^2\kappa_0\mu_0}}\\
a_2=&-i b_1\frac{f_{t,0}\sqrt{\kappa_0\mu_0}}{\sqrt{E_{x,0}^2+f_{t,0}^2\kappa_0\mu_0-c^2f_{x,0}^2\kappa_0\mu_0}}\\
a_3=&-i b_1\frac{c^2f_{x,0}\sqrt{\kappa_0\mu_0}}{\sqrt{E_{x,0}^2+f_{t,0}^2\kappa_0\mu_0-c^2f_{x,0}^2\kappa_0\mu_0}}\\
a_4=&-i b_1\frac{E_{x,0}}{\sqrt{\kappa_0\mu_0}\sqrt{E_{x,0}^2+f_{t,0}^2\kappa_0\mu_0-c^2f_{x,0}^2\kappa_0\mu_0}}\\
c_1=&c_2=c_3=c_4=0\\
d_1=&0\\
d_4=&\frac{d_2f_{t,0}+d_3f_{x,0}}{E_{x,0}}.
\end{align}
Now denoting
\begin{align}
D_0=&\frac{D}{i},
\end{align}
if we set $B_x(0)=B_{x,0}$, $B_y(0)=0$, $E_z(0)=0$ and $f_z(0)=0$, we obtain
\begin{align}
a_1=&b_1=\frac{1}{2}B_{x,0}\\
d_2=&d_3=0,
\end{align}
which gives
\begin{align}
B_x=&B_{x,0}\cos{D_0z}\\
B_y=&-\frac{B_{x,0}f_{t,0}\kappa\mu_0}{c D_0}\sin{D_0 z}\\
E_z=&-\frac{B_{x,0}c f_{x,0}\kappa\mu_0}{D_0}\sin{D_0 z}\\
f_z=&-\frac{B_{x,0}E_{x,0}\kappa}{c\kappa_0 D_0}\sin{D_0 z}
\end{align}
If we set $B_x(0)=B_{x,0}$, $B_y(0)=0$, $E_z(0)=0$ and $f_z(0)=f_{z,0}$, we obtain
\begin{align}
a_1=&\frac{1}{2}(B_{x,0}-\frac{i E_{x,0}f_{z,0}\sqrt{\kappa_0\mu_0}}{\sqrt{E_{x,0}^2+f_{t,0}^2-c^2f_{x,0}^2}})\\
b_1=&\frac{1}{2}(B_{x,0}+\frac{i E_{x,0}f_{z,0}\sqrt{\kappa_0\mu_0}}{\sqrt{E_{x,0}^2+f_{t,0}^2-c^2f_{x,0}^2}})\\
d_2=&-\frac{E_{x,0}f_{t,0}f_{z,0}\kappa_0\mu_0}{E_{x,0}^2+f_{t,0}^2-c^2f_{x,0}^2}\\
d_3=&-\frac{c^2E_{x,0}f_{x,0}f_{z,0}\kappa_0\mu_0}{E_{x,0}^2+f_{t,0}^2-c^2f_{x,0}^2}
\end{align}
which gives
\begin{align}
B_x=&B_{x,0}\cos{D_0z}+\frac{E_{x,0}f_{z,0}\kappa\mu_0}{c D_0}\sin{D_0 z}
\end{align}
\begin{align}
B_y=&\frac{f_{t,0}E_{x,0}f_{z,0}\kappa^2\mu_0^2}{c^2D_0^2}(-1+\cos{D_0 z})\nonumber\\
&-\frac{B_{x,0}f_{t,0}\kappa\mu_0}{c D_0}\sin{D_0 z}
\end{align}
\begin{align}
E_z=&\frac{f_{x,0}E_{x,0}f_{z,0}\kappa^2\mu_0^2}{D_0^2}(-1+\cos{D_0 z})\nonumber\\
&-\frac{B_{x,0}c f_{x,0}\kappa\mu_0}{D_0}\sin{D_0 z}
\end{align}
\begin{align}
f_z=&\frac{\kappa}{c^2D_0^2\kappa_0}[c D_0f_{z,0}\sqrt{\kappa_0\mu_0}+E_{x,0}^2f_{z,0}\kappa\mu_0(-1+\cos{D_0 z})\nonumber\\
&-B_{x,0}E_{x,0}c D_0\sin{D_0 z}].
\end{align}

\bibliographystyle{apsrev4-1}

\end{document}